\documentclass[12pt]{article}
\usepackage[dvips]{color}
\usepackage{amsmath}
\usepackage{graphicx}
\usepackage{subfigure}
\usepackage{epsfig}
\usepackage{amsmath}
\usepackage{cite}
\usepackage{color}
\usepackage{subfigure}
\usepackage{multirow}
\usepackage{float}

\begin{document}
\begin{center}
\Large{\bf Effective Potential and Topological Photon Spheres: A Novel Approach to Black Hole Parameter Classification }\\
 \small \vspace{0.5cm}
 {\bf  Mohammad Ali S. Afshar $^{\star}$\footnote {Email:~~~m.a.s.afshar@gmail.com}}, \quad
 {\bf Jafar Sadeghi $^{\star}$\footnote {Email:~~~pouriya@ipm.ir}}\\
\vspace{0.5cm}$^{\star}${Department of Physics, Faculty of Basic
Sciences,\\
University of Mazandaran
P. O. Box 47416-95447, Babolsar, Iran}\\
\small \vspace{0.5cm}
\end{center}
\begin{abstract}
In this paper, we base our analysis on the assumption that the existence of a photon sphere is an intrinsic feature of any ultra-compact gravitational structure with spherical symmetry. Utilizing the concept of a topological photon sphere, we categorize the behaviors of various gravitational models based on the structure of their photon spheres. This innovative approach enables us to define boundaries for black hole parameters, subsequently allowing us to classify the model as either a black hole or a naked singularity. Indeed, we will demonstrate that the presence of this interplay between the gravitational structure and the existence of a photon sphere is a unique advantage that can be utilized from both perspectives. Our observations indicate that a gravitational model typically exhibits the behavior of a horizonless structure (or a naked singularity) when a minimum effective potential (a stable photon sphere) appears within the studied spacetime region. Additionally, in this study, we tried to investigate the effect of this structure on the behavior of the photon sphere by choosing models that are affected by the Perfect Fluid Dark Matter (PFDM). Finally, by analyzing a model with multiple event horizons, we show that the proposed method remains applicable even in such scenarios.
\end{abstract}
\tableofcontents
\section{Introduction}
In the study of ultra-compact gravitational objects, the simplest and most fundamental step to distinguish between different configurations is to examine the metric function and subsequently identify the event horizon. The metric function is essentially the result of integrating equations based on diverse and complex initial conditions (such as energy conditions, geometric compatibility with general relativity equations, etc.). Any remaining weaknesses that might challenge the physics of the derived model are addressed by conjectures such as the Weak Cosmic Censorship Conjecture (WCCC), which posits the creation of an event horizon. This conjecture assumes that singularities emerging from solving Einstein's equations must be hidden behind an event horizon. This is because if singularities were observable from the rest of spacetime, causality would be compromised, and physics would lose its predictive power. It is important to note that the failure to meet the WCCC condition does not imply "non-existence" or "absence," but rather the "existence" of a spacetime that, although it has its own gravitational effects, our current knowledge is insufficient to decode it. Extensive studies are currently being conducted on the possibility and conditions for the emergence and observation of these regions. For instance, spacetime singularities formed during gravitational collapse can be observed by an external observer, and the observability of a spacetime singularity depends on the initial conditions of the collapsing matter \cite{1,2,3,4,5,6,7,8,9,10,11,12,13,14}. Therefore, it is crucial to know under what conditions our model behaves like a black hole and from what range it transitions to an ultra-compact object without an event horizon (naked singularity). However, naked singularity regions may also be involved under certain conditions in the results. Hence, their extent of influence must also be determined.\\\\ After constructing the initial metric function, the most significant factor influencing the location of the event horizon is the values assigned to the parameters of each theory used in constructing the metric function. In a simplified view, the permissible range of these parameters can be studied based on the roots of the metric function. Because all necessary conditions, including the geometric equations of general relativity and energy conditions, have been integrated within the system of equations and the result has manifested as a model characterized by a defined metric function. However, it is clear that this method will not be applicable in determining the dominance range of a naked singularity. Therefore, to gain a better understanding of the overall space surrounding an ultra-compact object, one must seek a potential function that provides a more comprehensive response to how the gravity of this structure affects the entire surrounding space. In pursuit of such a function, we know that the nature of gravitational structures, whether in their classical or quantum forms, necessitates the existence of stable and unstable circular orbits. Moreover, in gravity, the existence of such orbits around a black hole or an ultra-compact object has been well calculated both experimentally and observationally, as well as theoretically for various black hole models.For example, in a observational study, investigators documented the presence of a distinct dark shadow, which emerged due to the gravitational deflection of light by the colossal black hole located at the nucleus of the massive elliptical galaxy M87. This phenomenon conclusively substantiates the existence of photon rings encircling the central dense object \cite{15,16,17,18,19,20}. Also, theoretically, Cunha et al. showed during their studies that, in spherical symmetry, standard black holes, as well as other ultra-compact objects (that can be with or without event horizon) in general relativity, can have planar circular photon orbits. In such a way that the stable type causes instability, and its unstable type could determine the shadows of the black hole \cite{21}. Also, in a new paper it was shown in a geometrical proof that 'any spherically symmetric space-time that is asymptotically flat and has a horizon must contain a photon sphere, a spherical photon shell' \cite{22}. Therefore, considering the above characteristics, it seems that the idea of the photon sphere can be used for a more precise classification of the space surrounding the ultra-compact structures.\\ As, we have stated, the foundation of our work in this article is predicated on the necessity of the photon sphere for ultra-compact object structures. However, to approach the topological photon sphere historically, we commence our study of the photon sphere from the point where Cardoso and his colleagues have articulated: "The mere observation of an unstable light ring is strong evidence for the existence of black holes", or, "The existence of a stable light ring is thus an unavoidable feature of any ultra-compact star" \cite{23}. Then Cunha et al. showed during their studies that, in spherical symmetry, standard black holes, as well as other ultra-compact objects (that can be with or without event horizon) in general relativity, can have planar circular photon orbits.  In such a way that the stable type causes instability, and its unstable type could determine the shadows of the black hole \cite{22}. Finally, after Cunha et al. which extended their work to spherically symmetric 4D black holes\cite{26},” Shao-Wen Wei” used this idea and” Duan” mapping, and extended the discussion of the photon ring to photon sphere with a topological approach. He stated that there exists at least one standard photon sphere outside the black hole not only in asymptotically flat space-time, but also in asymptotically AdS and dS space-time \cite{24}.\\
In this article, we will use the "Wei" method to classify space based on the topological photon sphere behavior for different black holes. It means that, the topological current is non-zero only at the zero point of the vector field, which determines the location of the photon sphere. Therefore, one topological charge can be considered for each photon sphere. In the full exterior region, based on a specific range of parameters where the Total Topological Charge(TTC) is always equal to -1, the structure will be considered a black hole, and conversely. Moreover, on the same basis, for a naked singularity that has a vanishing topological charge, the TTC will be 0 \cite{24} or +1 \cite{25}. Ultimately, if neither of the above states occurs , these areas  will likely represent forbidden zones. Forbidden in the sense that both the metric function has no roots and the potential function shows no gravitational effect on photons.
An essential point to emphasize before the end of this section is that in this article, we utilize the effective potential which the notable characteristic of this potential is that it solely depends on the geometry and structure of spacetime, without reliance on the properties of the incoming particle. 
We will examine the existence of photon spheres in two scenarios: structures with an event horizon (black holes) and those without an event horizon (naked singularities). In addition to identifying the locations of these spheres and demonstrating the behavioral alignment of topological charge with the geometric behavior of the potential function, we will show how the permissible range for each parameter must be defined for the structure to mathematically manifest as either a black hole or a naked singularity.
\textit{It is crucial to note that our discussion is fundamentally based on the gravitational influence of the model on the motion of photons and massless particles, as well as the structure of null geodesics. Although the existence of these orbits is a necessary condition for the structure of an ultra-gravitational object, this does not imply that the obtained range necessarily holds completely physical meaning. This range will be most meaningful when interpreted alongside other physical conditions and constraints. Overall, this method can provide a comprehensive view of the behavior of ultra-gravitational structures and serve as an auxiliary equation, offering better insight into the impact of the effective components in the black hole dynamics of the model.}
\\All above information give us motivation to arrange the paper as follows. In section 2, we briefly review the mathematical and physical foundations of the work. The Duan’s topological mapping and the effective potential are introduced, and then we state the circular null geodesic conditions. And finally, we combine the obtained relations and deduce the used calculation. In Section 3, we first address the advantages and disadvantages of utilizing this method and proceed to study the effective potential with greater precision in terms of physical concepts and the alignment of its graphical results with the TTC outcomes. We will endeavor to articulate the physical significance of the effective potential graph.
In section 4,we will study and analyze different introduced models using this method and Finally, we have conclusions which are summarized in section 5.
\section{The topological path to the photon sphere}
 Solving the metric function (1) based on the specific parameters only leads to an interval in which  the radius of event horizon will remain real according to the desired parameters,
\begin{equation}\label{(1)}
\mathit{ds}^{2}=-\mathit{dt}^{2} f \! \left(r \right)+\frac{\mathit{dr}^{2}}{g \! \left(r \right)}+\left(d\theta^{2}+d\varphi^{2}
\sin \! \left(\theta \right)^{2}\right) h \! \left(r \right).
\end{equation}
The event horizon, a null one-sided hypersurface that acts as a causal boundary, ensuring that the WCCC holds. 
But there are other areas where system behavior is still important to us (for example, naked singularities, the hypothetical gravitational singularity without an  event horizon).
The meaning of the above statements will be that, there are regions where the behavior of the system is singular, and the metric function cannot  help us to identify that region. 
For this reason, we must turn to concepts that, with greater precision and over a broader area, display the gravitational behavior of the model under study. Consequently, we explore the photon ring and photon sphere.
The photon sphere is a null ring that is the lower bound for any stable orbit, showing the extreme bending of light rays in a very strong gravity. 
It has two types: unstable, where small perturbations make the photons either escape or fall into the black hole, and  stable, where the opposite happens. 
The unstable type is useful for determining the black hole  shadows, while the stable type causes spacetime instability\cite{27}.\\  
Several methods can be used to study the photon sphere that among the different  possible methods, in this work we considered the ”Wei” topological method \cite{24}. If we want to  analyze the foundations of the work of this will be combination of basic concepts such as Poincaré section \cite{28}, Duan’s topological current mapping theory and the necessary conditions for the existence of null geodesics.\\
In 1984, Yishi Duane conducted a pivotal analysis of the intrinsic structure of conserved topological currents within the SU(2) gauge theory. During this examination, Duane introduced the concept of topological flow associated with point-like systems akin to particles. This foundational work laid the groundwork for subsequent discussions on various forms of topological currents.\cite{29}.
In the first step, we consider a general vector field as  $\phi$  which can be decomposed into two components, $\phi^r$ and $\phi^\theta$,
\begin{equation}\label{1}
\phi=(\phi^{r}, \phi^{\theta}),
\end{equation}
 also here, we can  rewrite the vector as $\phi=||\phi||e^{i\Theta}$, where $||\phi||=\sqrt{\phi^a\phi^a}$, or $\phi = \phi^r + i\phi^\theta$.
 Based on this, the normalized vector is defined as,
 \begin{equation}\label{2}
n^a=\frac{\phi^a}{||\phi||},
\end{equation}
 where $a=1,2$  and  $(\phi^1=\phi^r)$ , $(\phi^2=\phi^\theta)$.
Now we introduce our antisymmetric superpotential as follows\cite{24,29},
 \begin{center}
 $\Upsilon^{\mu\nu}=\frac{1}{2\pi} \epsilon^{\mu\nu\rho} \epsilon_{ab}n^a\partial_\rho n^b,\hspace{0.3cm}\mu,\nu,\rho=0,1,2$,
\end{center}
and the topological current will be as,
\begin{center}
 $j^{\mu}=\partial_{\nu}\Upsilon^{\mu\nu}=\frac{1}{2\pi}\epsilon^{\mu\nu\rho}\epsilon_{ab}\partial_{\nu}n^a \partial_\rho n^b$.
\end{center}
Based on this, Noether's current and charges at the given $\Omega$ will be,
\begin{center}
 $\partial_\nu j^\nu=0$
 \end{center}
and
\begin{equation}\label{3}
Q=\int_{\Omega}j^0d^{2}x,
\end{equation}
where $j^0$ is the charge density.
By replacing $\phi$ instead of $n$ and using the Jacobi tensor, we will arrive at,
\begin{equation}\label{(4)}
j^{\mu}=\frac{J^{\mu} \! \left(X \right) \ln \! \left({||\phi||}\right) \Delta_{\phi^{a}}}{2 \pi},
\end{equation}
where $X=\frac{\phi}{x}$. By using  two-dimensional Laplacian Green function in $\phi-mapping$ space, we have,
\begin{equation}\label{(5)}
\ln \! \left({||\phi||}\right) \Delta_{\phi^{a}}=2 \delta \! \left(\phi \right) \pi,
\end{equation}
and the topological current  will be,
\begin{equation}\label{(6)}
j^{\mu}=J^{\mu} \! \left(X \right) \delta^{2} \! \left(\phi \right).
\end{equation}
From the properties of $\delta$, it is clear that $j^{\mu}$ is non-zero only at the zero points of $\phi^{a}$, and this is exactly what we need to continue the discussion.
Finally, by using the above relations and inserting them in relation (3), the topological charge become,
\begin{equation}\label{7}
Q=\int_{\Omega}J^{0} \! \left(X \right) \delta^{2} \! \left(\phi \right)d^{2}x,
\end{equation}
once again and this time around the topological charge Q, according to the characteristic of the $\delta$ function, it can be said that the charge is non-zero only at the zero point of $\phi$.
This results lead us  how to use and  calculate the photon sphere in the near future.\\
Now, we are going to investigate the photon sphere and some null geodesics with static and  spherical symmetry background (1),the Lagrangian takes the form \cite{23}:
\begin{equation*}\label{(0)}
\mathcal{L} ={x^{\cdot}}^{\nu} {x^{\cdot}}^{\mu} g_{\mu \nu} ={t^{\cdot}}^{2} f \! \left(r \right)-\frac{{r^{\cdot}}^{2}}{g \! \left(r \right)}-{\varphi^{\cdot}}^{2} h \! \left(r \right),
\end{equation*}
where the dot denotes a derivative with respect to an affine parameter. Now we could derive the four-momentum from this Lagrangian, we have:
\begin{equation*}\label{(0)}
p_{\mu}=\frac{\partial}{\partial {x^{\cdot}}^{\mu}}\mathcal{L}={x^{\cdot}}^{\nu} g_{\mu \nu},  
\end{equation*}
or more precisley
\begin{equation*}\label{(0)}
\begin{split}
& p_{t}=t^{\cdot} f \! \left(r \right)\equiv E,\\
& p_{\varphi}=-\varphi^{\cdot} h \! \left(r \right)\equiv -L,\\
& p_{r}=-\frac{r^{\cdot}}{g \! \left(r \right)}\\.
\end{split}
\end{equation*}
Where E  and L  respectively interpreted as the energy and angular momentum. As is evident from the Lagrange equation above, this function is independent of t and $\phi$, so $p_{t}$ and $p_{\phi}$ will be the two integrals of motion, namely:
\begin{equation*}\label{(0)}
\begin{split}
& \varphi^{\cdot}=\frac{L}{h \! \left(r \right)}\\
& t^{\cdot}=\frac{E}{f \! \left(r \right)}.\\
\end{split}
\end{equation*}
Now for the general form of Hamiltonian we have:
\begin{equation*}\label{(0)}
\Pi ={x^{\cdot}}^{\mu} p_{\mu}-\mathcal{L}= \Delta,
\end{equation*}
where $\Delta = 1 , 0$ for time-like and null geodesics.
Here, we consider four-momentum and Hamiltonian condition for null geodesics,
\begin{equation}\label{(9)}
\Pi =\frac{p^{\nu} p^{\mu} \textit{g}_{\mu ,\nu}}{2}=0,
\end{equation}
and the radial component of the null geodesic equations  will be as \cite{24},
\begin{equation}\label{10}
\begin{split}
\dot{r}^{2}+V_{eff}=0,
\end{split}
\end{equation}
and
\begin{equation}\label{11}
\begin{split}
V_{eff}=g(r)\bigg(\frac{L^2}{h(r)}-\frac{E^2}{f(r)}\bigg),
\end{split}
\end{equation}
 where $E$ and $L$ represent the photon’s energy and the total angular momentum, respectively.
 A circular null geodesic occurs at an extremum of the effective potential $V_{eff}$(r), which is given by,
 \begin{equation}\label{12}
\begin{split}
V_{eff}=0,\hspace{1cm}\partial_{r}V_{eff}=0.
\end{split}
\end{equation}
These local extrema of the effective potential  are equivalent to unstable and stable circular null geodesics, will correspond to a photon sphere and an anti-photon sphere.
By Considering the equations(12) at the same time, we have  following expression,
\begin{equation}\label{13}
\begin{split}
\bigg(\frac{f(r)}{h(r)}\bigg)^{'}_{r=r_{ps}}=0,
\end{split}
\end{equation}
where prime is the derivative with respect to $r.$
By rewriting(13), we will have,
\begin{equation}\label{14}
\begin{split}
f(r)h(r)'-f(r)'h(r)=0.
\end{split}
\end{equation}
The first term at the horizon will be disappear, but this case the second term usually remains non-zero, it means that $r_{ps}$ and $r_{h}$ do not coincide.
According to the results and concepts of the above discussion, we are now in a place where we can start investigating the topological characteristics of the photon sphere.\\So,here we Start this work by introducing a regular potential\cite{6},
\begin{equation}\label{15}
\begin{split}
H(r, \theta)=\sqrt{\frac{-g_{tt}}{g_{\varphi\varphi}}}=\frac{1}{\sin\theta}\bigg(\frac{f(r)}{h(r)}\bigg)^{1/2},
\end{split}
\end{equation}
the discussion of potential will allow us to search for the radius of our photon sphere at,
\begin{center}
  $\partial_{r}H=0$,
 \end{center}
 so, we can use a vector field $\phi=(\phi^r,\phi^\theta)$,
\begin{equation}\label{16}
\begin{split}
&\phi^r=\frac{\partial_rH}{\sqrt{g_{rr}}}=\sqrt{g(r)}\partial_{r}H,\\
&\phi^\theta=\frac{\partial_\theta H}{\sqrt{g_{\theta\theta}}}=\frac{\partial_\theta H}{\sqrt{h(r)}}.
\end{split}
\end{equation}
With the above definition $\phi$, and recalling the relationships of the Duan’s section, we now define the current and charge (3) for this new potential.\\
Furthermore, in light of relation (7) and the distinctive properties of the Dirac delta function, it can be inferred that the charges will manifest as non-zero exclusively at the loci where $\phi$ vanishes. Precisely at these junctures, the photon sphere is situated, thereby enabling the assignment of a topological charge Q to each photon sphere. Pursuant to equation (7), upon considering $\Omega$ as a manifold encompassing a singular zero point, subsequent analysis reveals that the charge Q is precisely commensurate with the winding number. Let $C_i$ denote a closed, smooth, and positively oriented curve that encapsulates solely the $i_{th}$ zero point of $\phi$, while all other zero points lie external to it." So the winding number can be calculated by the following formula,
\begin{equation}\label{17}
\omega_i=\frac{1}{2\pi}\oint_{C_{i}}d\Lambda,
\end{equation}
where $\Lambda$ is
\begin{equation}\label{18}
\Lambda=\frac{\phi^2}{\phi^1},
\end{equation}
then the total charge will be,
\begin{equation}\label{19}
Q=\sum_{i}\omega_{i}.
\end{equation}
In conclusion, when the closed curve encompasses a zero point, the topological charge Q is precisely equivalent to the winding number. In instances where the curve encloses multiple zero points, Q will be the aggregate of the winding numbers corresponding to each zero point. Conversely, should the curve circumscribe no zero points, the resultant charge must invariably be zero.\\
\section{Explanation of topological charges based on properties of effective potential}
Before delving into further details, perhaps the most important question to address is: What insights can the study of different photon sphere structures provide regarding the nature of black holes? 
It is important to state that the stability/instability of photon orbits around black holes plays a crucial role in shaping the observed black hole shadow, and can also provide insights into the presence of naked singularities. 
The shadow of a black hole is essentially the silhouette created by the event horizon against the backdrop of light from the accretion disk or other sources. The shape and size of this shadow are influenced by the photon orbits, which are paths that light can take around the black hole. The EHT’s observations of M87* and Sgr A* provide invaluable data on black hole shadows. These observations confirm the predictions of stable orbits affecting the shape and size of the shadow. The amount of light that visibly circles the black hole and contributes to its shadow can show insights into the structure surrounding it, potentially revealing the presence of an accretion disk, which influences photon paths \cite{29.1}.
For a non-rotating black hole, the shadow is nearly circular. However, for a rotating black hole, the shadow becomes distorted due to frame-dragging effects, where space-time itself is twisted by the black hole's rotation \cite{29.2}. The size of the shadow is determined by the radius of the photon sphere. A larger photon sphere results in a larger shadow.  
If photon orbits were stable, the shadow would be more sharply defined, as photons would orbit the black hole for longer periods before either escaping or being captured \cite{29.3}. 
In reality, the unstable orbits, leading to a more diffuse shadow boundary. This instability causes the photons to either spiral into the black hole or escape, contributing to the characteristic "fuzzy" edge of the shadow\cite{29.4}.
But the story about the naked singularity is completely different. Naked singularities, if they exist, would present a very different observational signature compared to black holes The lack of an event horizon means that photons could potentially orbit much closer to the singularity, leading to more extreme bending of light and potentially observable differences in the light patterns. 
This would lead to unique gravitational lensing effects, where light from background stars or other objects would be bent in unusual ways. 
In these cases, instead of well-defined shadows, one might expect unique visual signatures - such as anomalously bright phenomena formed by gravitational lensing effects or different patterns of light emissions due to their unstable nature \cite{29.5}.\\ 
In his paper, ’Wei’ demonstrated through the examination of winding numbers and the calculation of total charges for a model that whenever the TTC is -1, the structure will exhibit black hole behavior, and conversely, if the TTC is 0, the structure will be in the form of a naked singularity \cite{24}. We have extended this work to show that in addition to the value of 0, the TTC value of +1 can also appear as a naked singularity \cite{25}. In addition to utilizing the aforementioned analysis and topological charge analysis to classify the behavior of the studied structure and determine its parameter ranges, in this paper we also examine the geometric behavior of the effective potential (15), which shows precise alignment with the topological charges. A significant and distinct aspect of this potential, compared to traditional potentials used for studying photon spheres and null geodesics, is that it is independent of the energy and angular momentum of the incoming particle. This implies that this potential is an inherent property of the spacetime structure. Our observations in multiple cases \cite{25.1,25.2,25.3} indicate that when the studied system exhibits a topological charge of -1, it always possesses the event horizon and displays black hole behavior. In such a scenario, the potential function reveals a dominant maximum outside the event horizon, indicating instability at the photon sphere location. Conversely, when the ultra-compact object under study exhibits a topological charge of 0 or +1, the system generally lacks an event horizon and takes the form of a naked singularity. In this case, the potential function reveals a dominant minimum alongside the maximum, indicating the presence of a stable photon sphere.
It is worth noting an important and intriguing point here. From a classical perspective, the existence of a minimum energy state (stable condition) alongside a maximum energy state (unstable condition) implies that the system will naturally and spontaneously move towards the stable minimum. Consequently, it appears that the emergence of a stable photon sphere in proximity to an unstable photon sphere will cause any photon capable of leaving the unstable maximum to be absorbed by the nearest stable minimum. This suggests that the space-time in question should be considered "off" from the perspective of observation and information reception.\\Before dealing with the main models and for a better understanding of each of the topological states stated above, we will see an example separately below.\\
 \begin{center}
\textbf{Case I: TTC =-1}
\end{center}
Typically, in such a scenario, one or several zero points emerge outside the event horizon in a manner that maintains the total charges at -1 and the structure with the unstable photon sphere remained in the form of a black hole.\\
As observed in Fig (1a), this black hole possesses a single zero point outside its event horizon, resulting in a TTC of -1, in accordance with reference \cite{24}. 
In Fig (1b), we have plotted the potential H(r). It is evident that this potential has a global maximum, which upon closer inspection of Fig (1c), is precisely located at the zero point or in other words, at the same location as the photon sphere. 
Given that this point represents a global maximum, any photon trapped within this potential will abandon this peak at the slightest disturbance, in pursuit of a state with lower energy. This signifies that we possess an unstable photon sphere, precisely the phenomenon of observational interest.
\begin{figure}[H]
 \begin{center}
 \subfigure[]{
 \includegraphics[height=6.5cm,width=8cm]{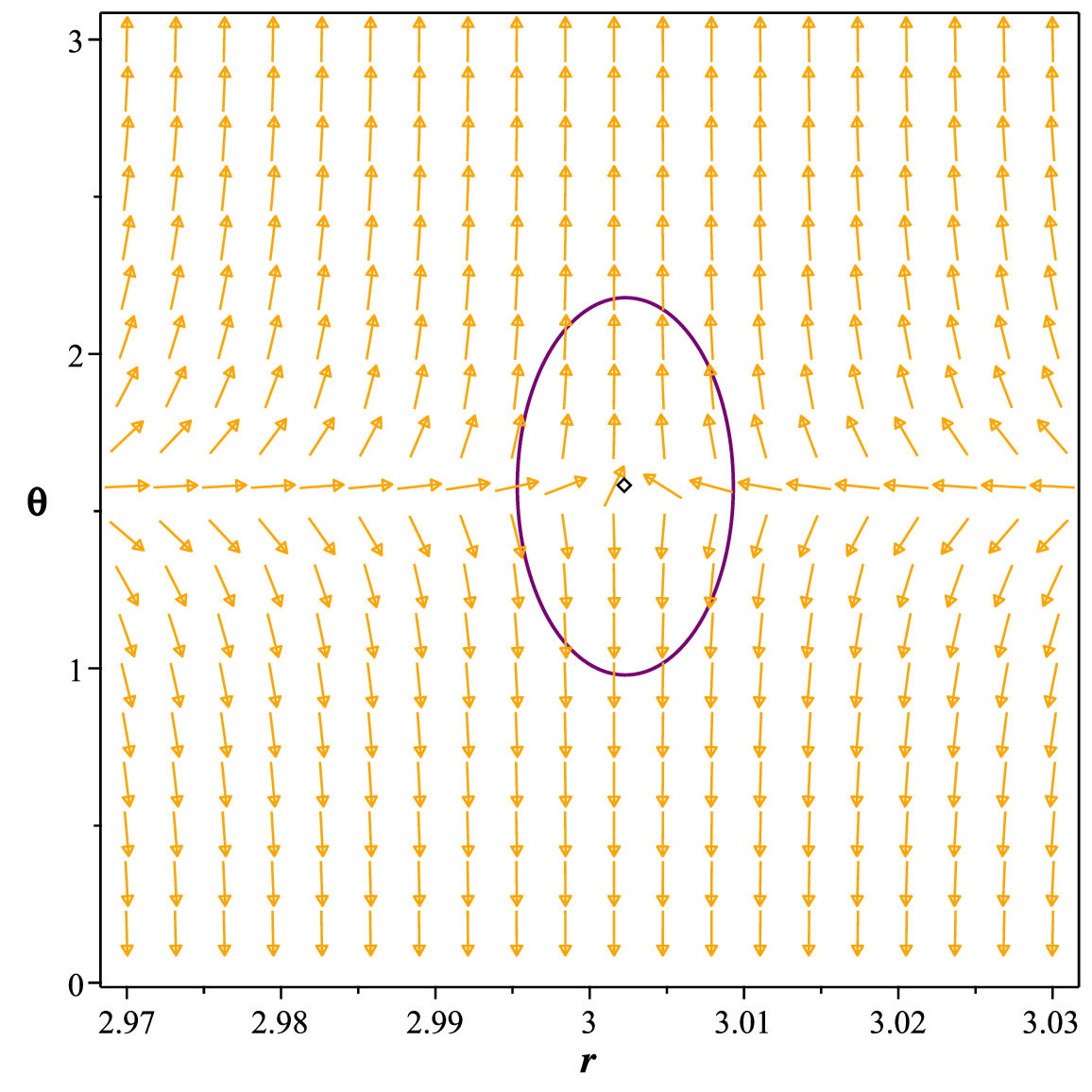}
 \label{1a}}
 \subfigure[]{
 \includegraphics[height=6.5cm,width=8cm]{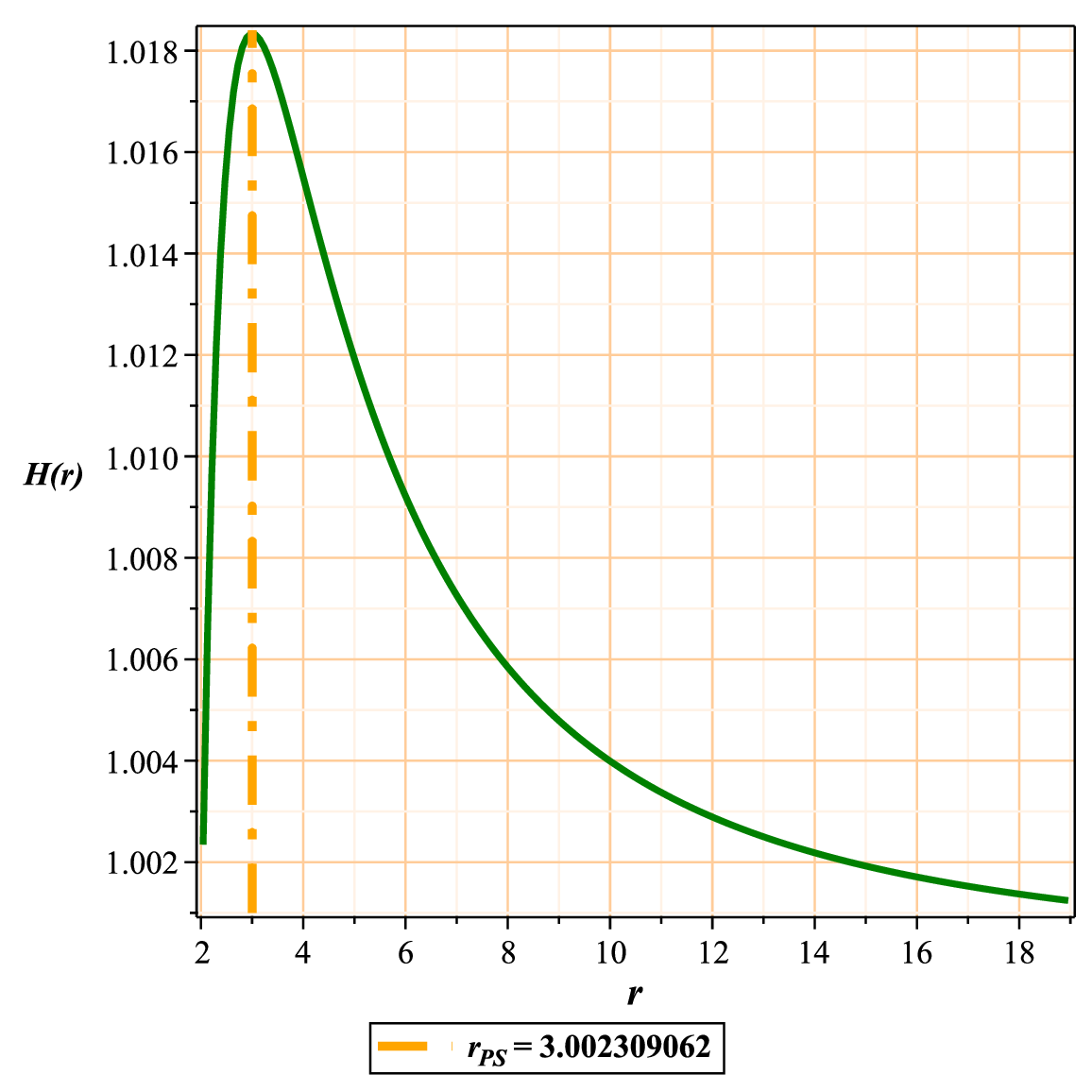}
 \label{1b}}
   \caption{\small{Fig (1a):The normal vector field $n$ in the $(r-\theta)$ plane. The photon sphere is located at $ (r,\theta)=(3.002309062,1.57)$ with respect to $(g = -0.21834, m = 1, l = 1)$  , (1b): the topological potential H(r) for regular Hayward AdS black hole model }}
 \label{5}
\end{center}
\end{figure}
\begin{center}
\textbf{Case II: TTC = 0 or +1}
\end{center}
From an energy perspective, these two states occur when one or more local or global minima appear in the studied spacetime. For the topological charge 0 case, as observed in Figure (2b), a minimum—a stable photon sphere—at ( r = 1.389900878 ) can be clearly seen adjacent to a maximum—an unstable photon sphere. It is important to note a few points here. Firstly, this state typically occurs when, from the metric function's viewpoint, we are in a rootless region, without a horizon, or in a space termed as a naked singularity. In other words, although these minima have always existed, even in the black hole state, but their influence on the final outcome only becomes apparent when the event horizon has disappeared, which for an ultra-compact object equates to a naked singularity.
The second point to consider is whether the structural influence in the form of a naked singularity is unlimited. The answer depends on the model under study; sometimes, parametric conditions arise where neither the metric function nor the potential function has a solution for the studied interval. In other words, beyond that parametric range, neither the metric function nor the potential function practically provides a solution for the space, which we refer to as the forbidden region.
\begin{figure}[H]
 \begin{center}
 \subfigure[]{
 \includegraphics[height=6.5cm,width=8cm]{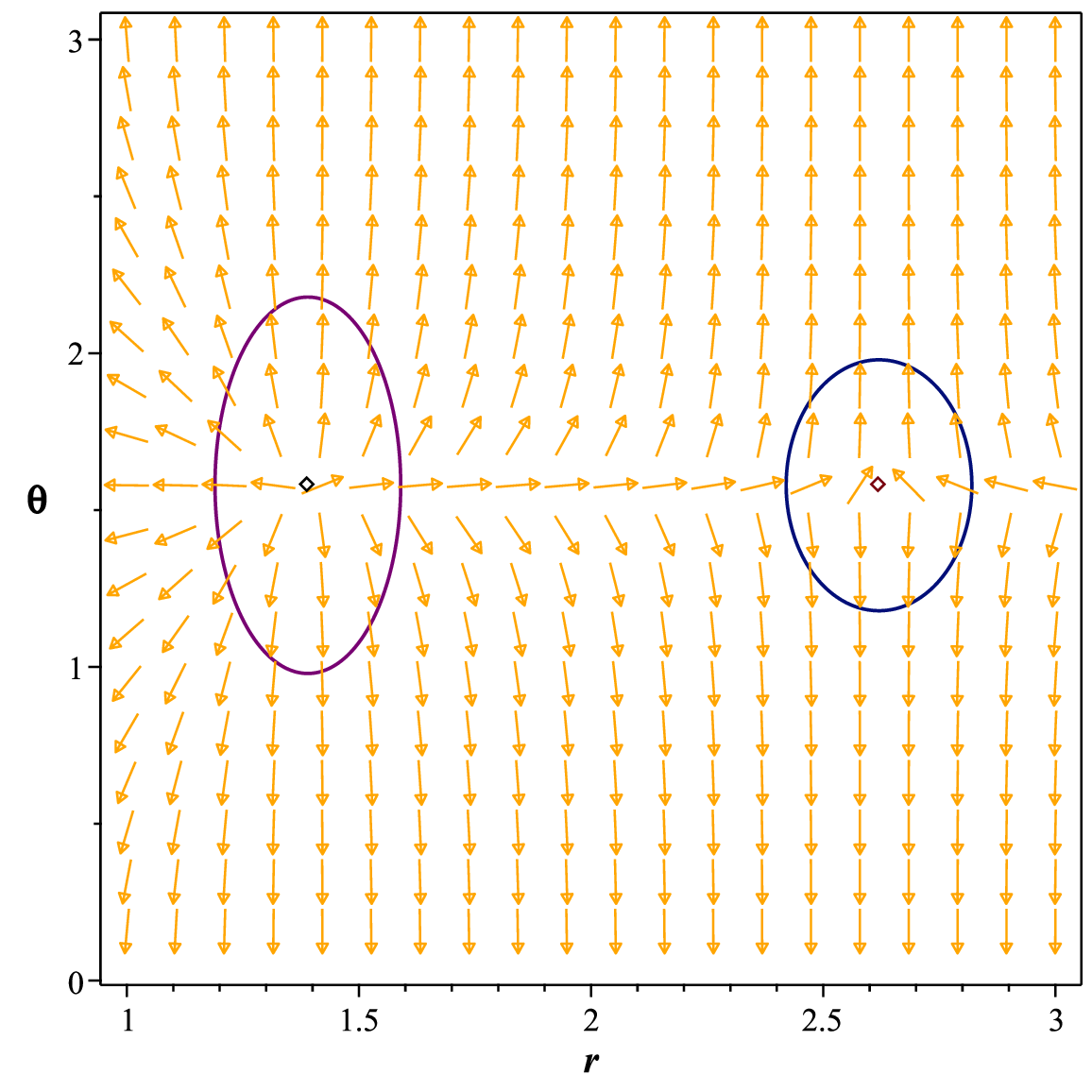}
 \label{2a}}
 \subfigure[]{
 \includegraphics[height=6.5cm,width=8cm]{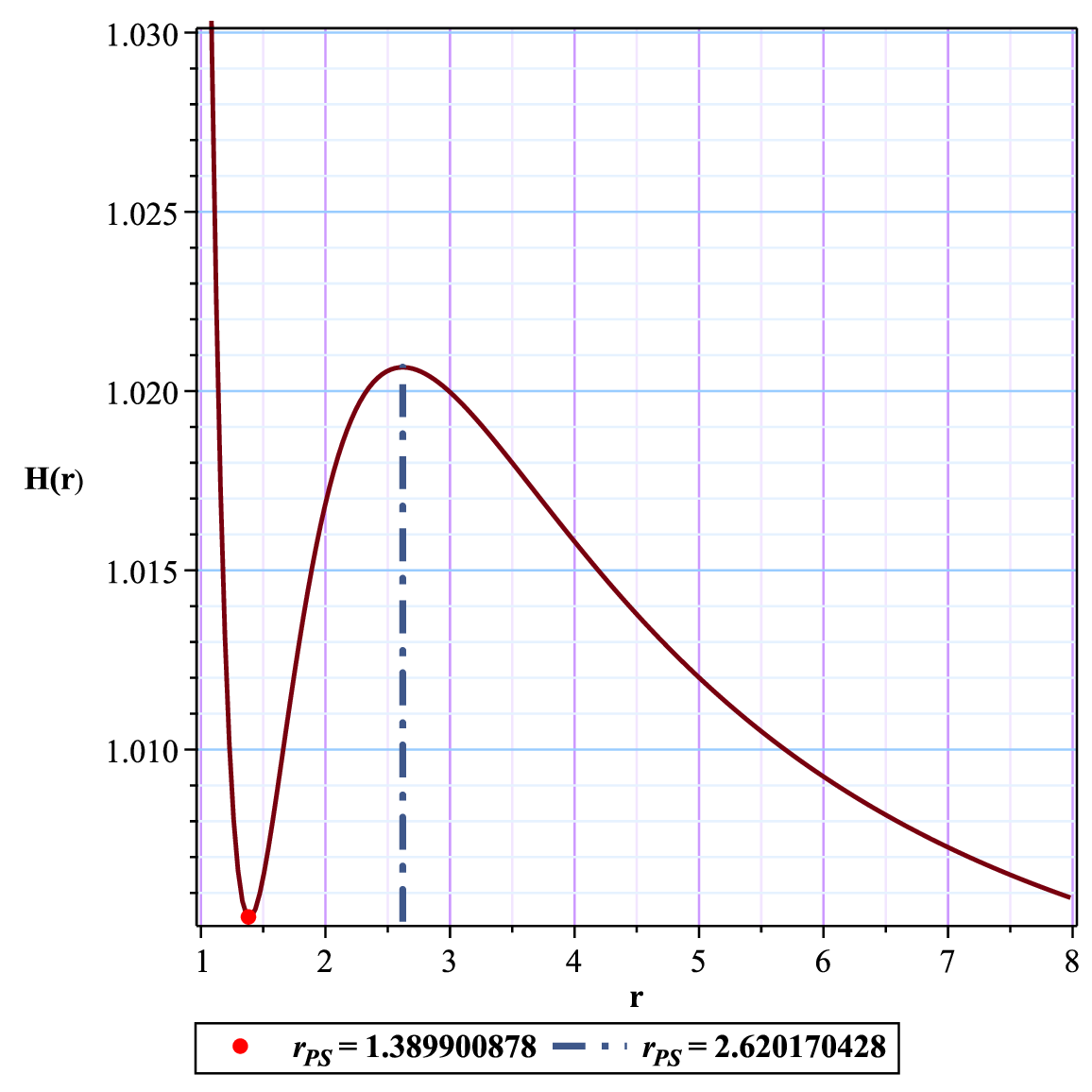}
 \label{2b}}
 
   \caption{\small{Fig (2a):The normal vector field $n$ in the $(r-\theta)$ plane. The photon spheres are located at $ (r,\theta)=(1.389900878,1.57)$ and $ (r,\theta)=(2.620170428,1.57)$   with respect to $( g=1.08, m=1,l=1 )$  , (2b): the topological potential H(r) for regular Hayward AdS black hole model }}
 \label{2}
\end{center}
\end{figure}
In Fig(3), we encounter a scenario with a total topological charge (TTC) of +1.In this case, more than one minimum appears in the studied space-time.
\begin{figure}[H]
 \begin{center}
 \subfigure[]{
 \includegraphics[height=6.5cm,width=8cm]{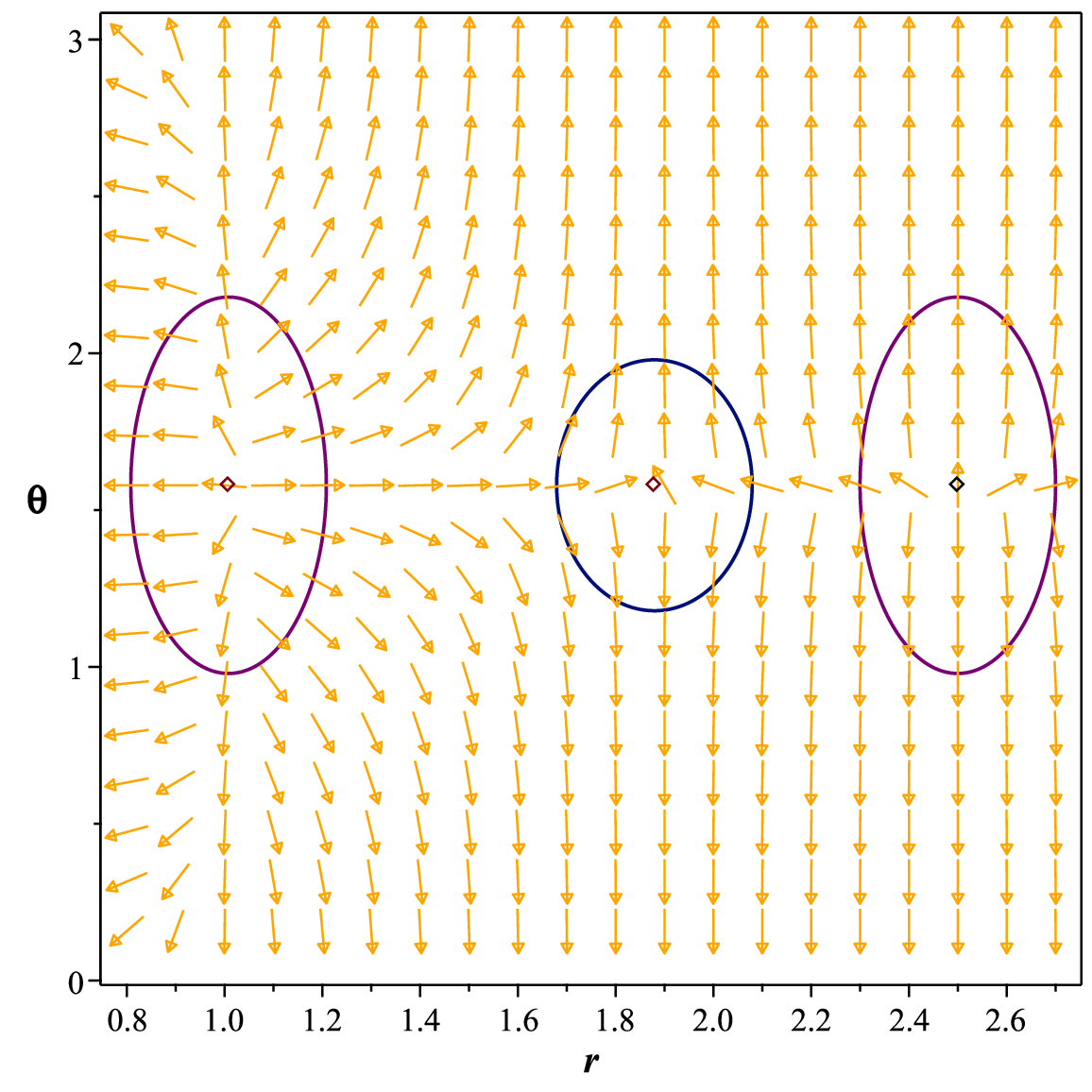}
 \label{3a}}
 \subfigure[]{
 \includegraphics[height=6.5cm,width=8cm]{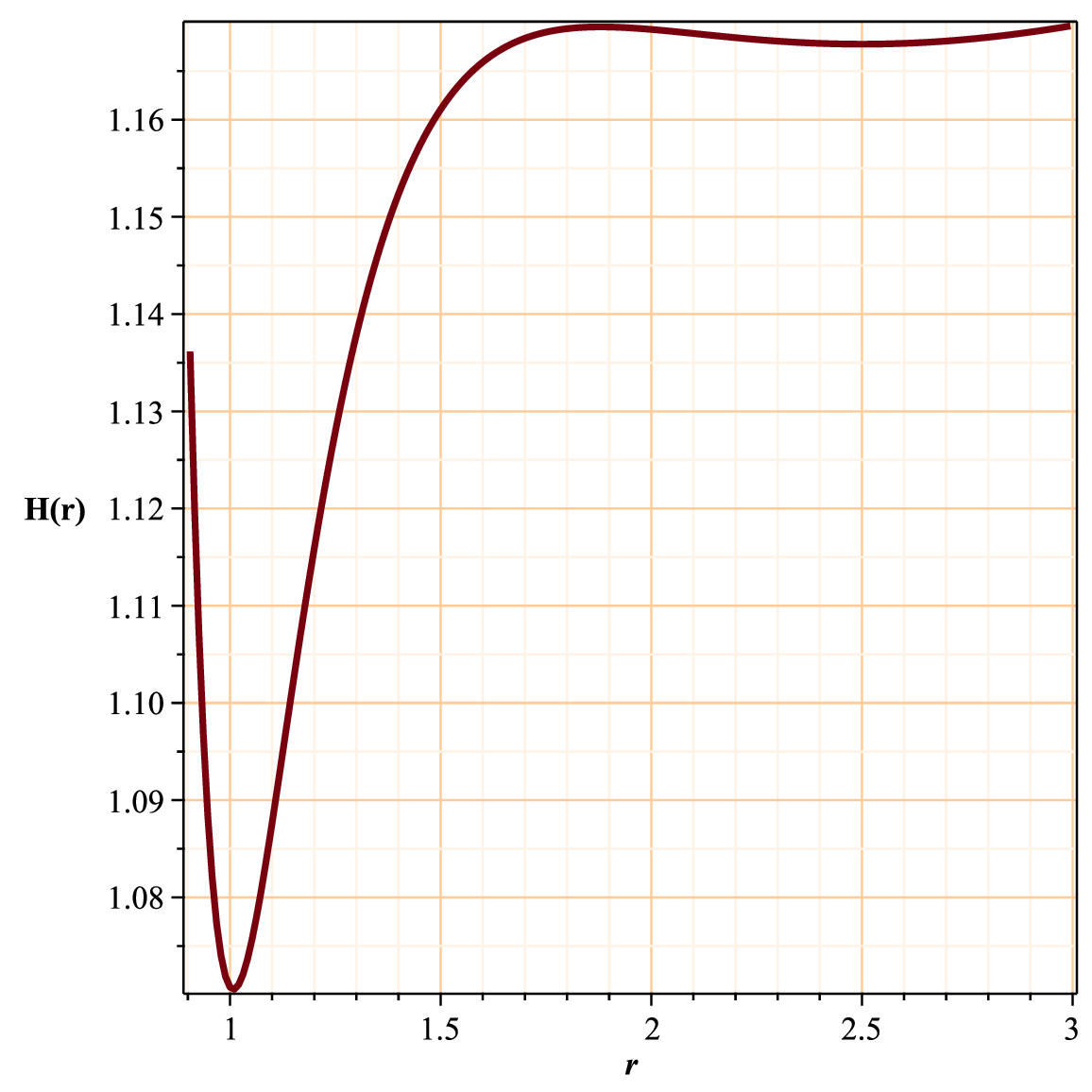}
 \label{3b}}
 \subfigure[]{
 \includegraphics[height=6.5cm,width=8cm]{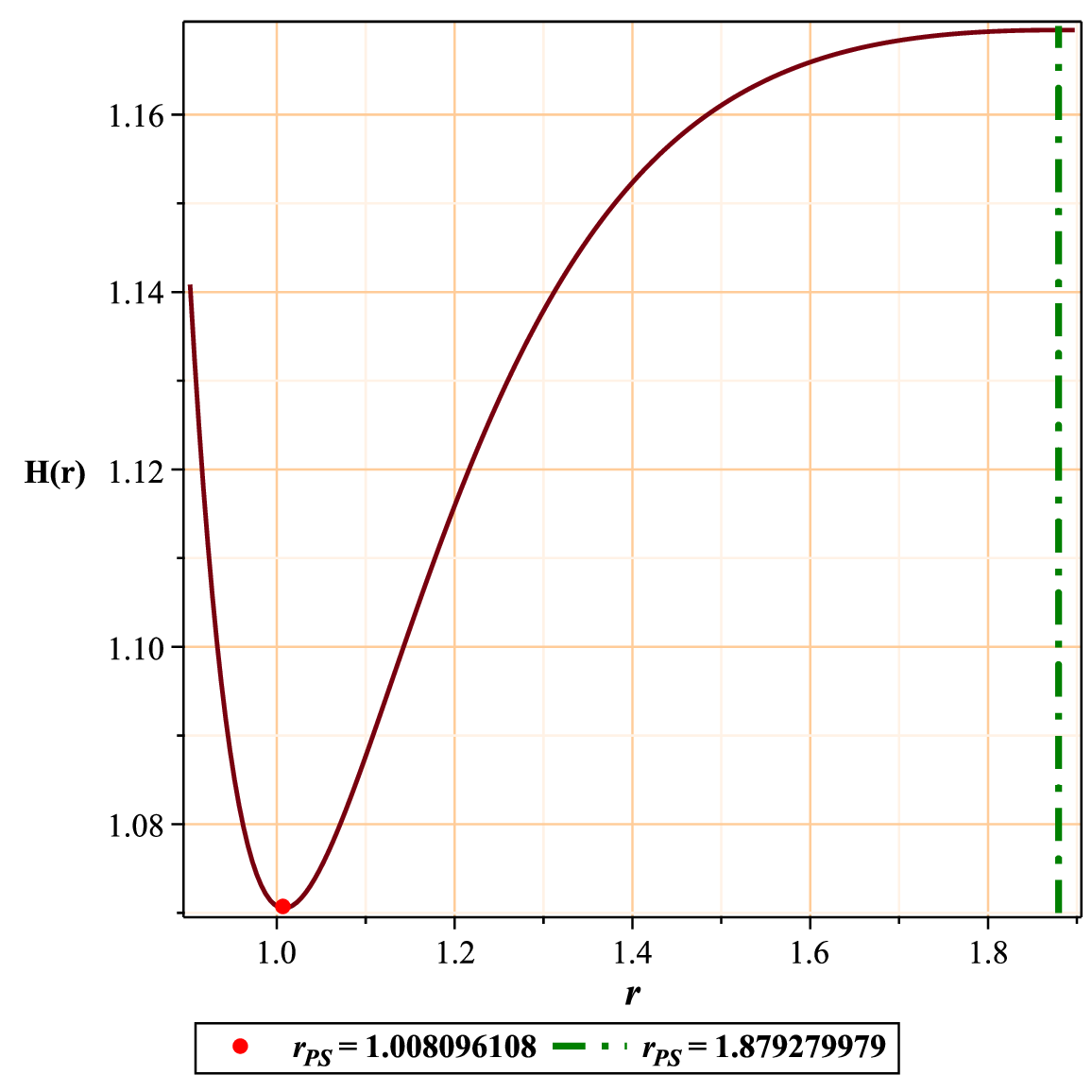}
 \label{3c}}
 \subfigure[]{
 \includegraphics[height=7cm,width=8cm]{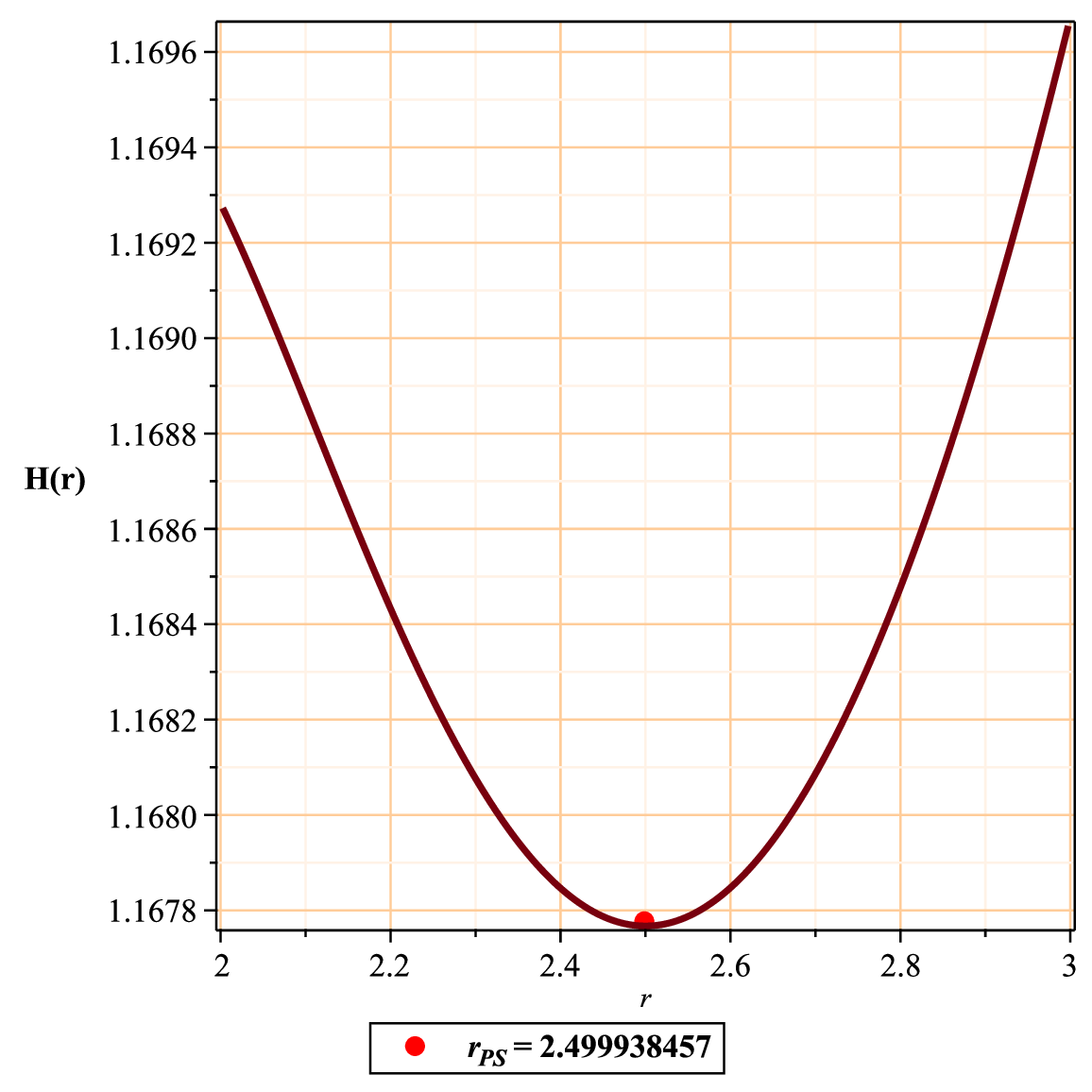}
 \label{3d}}
   \caption{\small{Fig (3a):The PSs are located at $(r-\theta)$ = (2.499938457, 1.57),$(r-\theta)$ = (1.879279979, 1.57), $(r-\theta)$ =  (1.008096108, 1.57) with respect to $(q = 0.9, z = 1.1, M = 1.6, \theta = 0)$ , (3b): the topological potential H(r) for Hyperscaling violating  model  , (3c),(3d): Enlarged details of diagram (3b)  }}
 \label{3}
\end{center}
\end{figure}
\section{Effective potential and topological photon sphere as a tool for classification}
Before delving into the selected models, it is beneficial to briefly compare the advantages and disadvantages of this method relative to the traditional approach. 
In the traditional approach to studying photon spheres, it is essential to extract the Lagrangian from the action and subsequently construct the Hamiltonian. Once the Hamiltonian is determined, the effective potential can be constructed, which allows for the study of the photon sphere. This potential is a function of the energy and angular momentum of the particle.\\
One notable advantage of this method is the direct use of the metric function to construct the potential, bypassing all the aforementioned steps. Additionally, the resulting potential is independent of the parameters of the incoming particle and is solely a function of the surrounding spacetime geometry. Using the equatorial plane for the vector field ($\phi$), similar to the Poincaré plane, and the resulting dimensional reduction, is another significant advantage. Ultimately, the ability to study a broader range of the space surrounding an ultra-compact object is a fundamental aspect that, for the first time, leads to the classification of different spacetime regions based on this method. Finally, this method allows for the introduction of a practical concept, which we term the possible radius limit for the photon sphere.\\ $R_{PLPS}$: Considering the permissible parameter range to maintain the black hole structure, we define the minimum or maximum possible radius for the appearance of an unstable photon sphere as $R_{PLPS}$.\\ However, there are fundamental drawbacks that must be considered. Since the vector field ($\phi$) must be separable with respect to spatial coordinates, only diagonal metrics with spherical symmetry can be used. Given that most models of interest have spherical structures and the rest can often be expressed diagonally using the C-metric, this limitation can be somewhat overlooked. Another issue is that since the Poincaré plane is used for dimensional reduction, this method may not be applicable to models with fewer than four dimensions. Finally, as this is a nascent approach, there may be weaknesses that become apparent over time. Interestingly, the simplifications, computational power, and novelty of this method and mathematical model for studying symmetric potential functions have garnered such attention that the mathematical structure of the model has been extended to thermodynamics. Consequently, extensive studies based on the mathematical model of this method have been conducted on the phase transitions of black hole models \cite{29.6,29.7,29.8,29.81,29.82,29.9,29.10,29.11,29.12,30,31,32,33,34,35,36,37,38}.
\subsection{Photon Sphere and perfect fluid dark matter black hole(PFDM)}
Cosmological calculations show that the surrounding universe contains
{$ \displaystyle  \mathit{\%\!73}  $} dark energy, {$ \displaystyle  \mathit{\%\!23}  $} dark matter, and the rest consists of baryonic matter\cite{39,40,41}.
This issue has been confirmed, to some extent, by baryonic sound oscillations, cosmic microwave background, weak lensing and other possible methods.
Also, astrophysical observations show that massive black holes, surrounded by huge halos of dark matter, are located in the centers of giant spiral and elliptical galaxies.
These evidences can be acceptable reasons for the fact that we should pay more attention to black hole solutions in the presence of dark matter and dark energy\cite{42,43,44,45,46}.
For this purpose, we chose the model  four-dimensional PFDM\cite{47},the metric for such black hole is
\begin{figure}[H]
 \begin{center}
 \subfigure[]{
 \includegraphics[height=6.5cm,width=8cm]{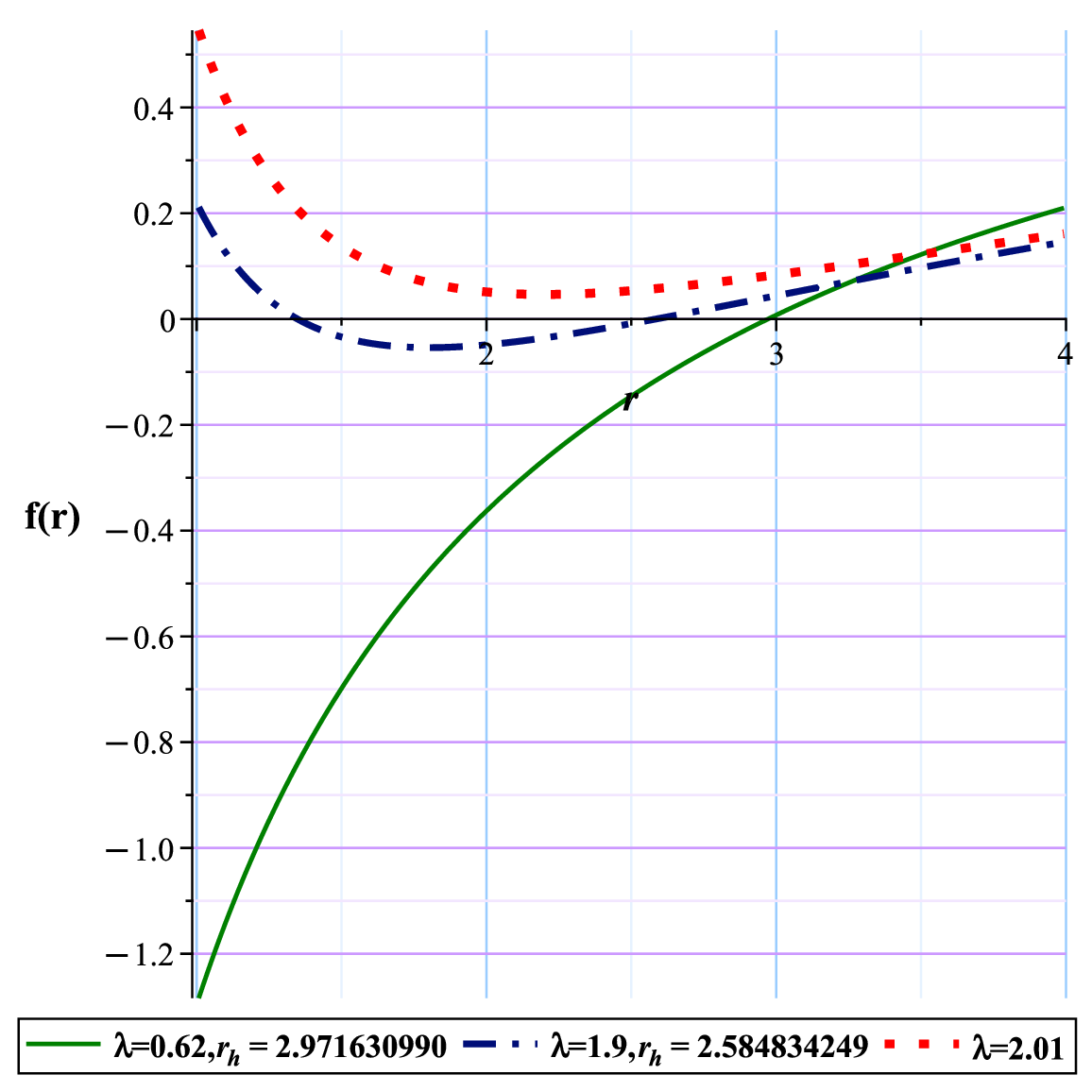}
 \label{4a}}
 \caption{\small{Metric function with different $\lambda$ for PFDM black hole model  }}
 \label{4}
\end{center}
\end{figure}
\begin{equation}\label{(20)}
f =1-\frac{2 M}{r}-\frac{\lambda  \ln \! \left(\frac{r}{\lambda}\right)}{r},
\end{equation}
where M is mass and $ \lambda$ is the parameter of intensity of PFDM.  Also, we have,
\begin{equation}\label{(21)}
f \! \left(r \right)=g \! \left(r \right),
\end{equation}
\begin{equation}\label{(22)}
h \! \left(r \right)=r,
\end{equation}
From Eq(15) for potential and with respect to above equations we have:
\begin{equation}\label{(20)}
H =\frac{\sqrt{1-\frac{2 M}{r}-\frac{\lambda  \ln \left(\frac{r}{\lambda}\right)}{r}}}{\sin \! \left(\theta \right) r}.
\end{equation}
With respect to(21),(22) and from equation (16), we will have:
\begin{equation}\label{(20)}
\phi^{r}=\frac{\left(3 \ln \! \left(\frac{r}{\lambda}\right) \lambda +6 M -2 r -\lambda \right) \csc \! \left(\theta \right)}{2 r^{3}},
\end{equation}
\begin{equation}\label{(20)}
\phi^{\theta}=-\frac{\sqrt{1-\frac{2 M}{r}-\frac{\lambda  \ln \left(\frac{r}{\lambda}\right)}{r}}\, \cos \! \left(\theta \right)}{\sin \! \left(\theta \right)^{2} r^{2}}.
\end{equation}
\begin{center}
\textbf{Case I: TTC =-1}
\end{center}
\begin{figure}[H]
 \begin{center}
 \subfigure[]{
 \includegraphics[height=6.5cm,width=8cm]{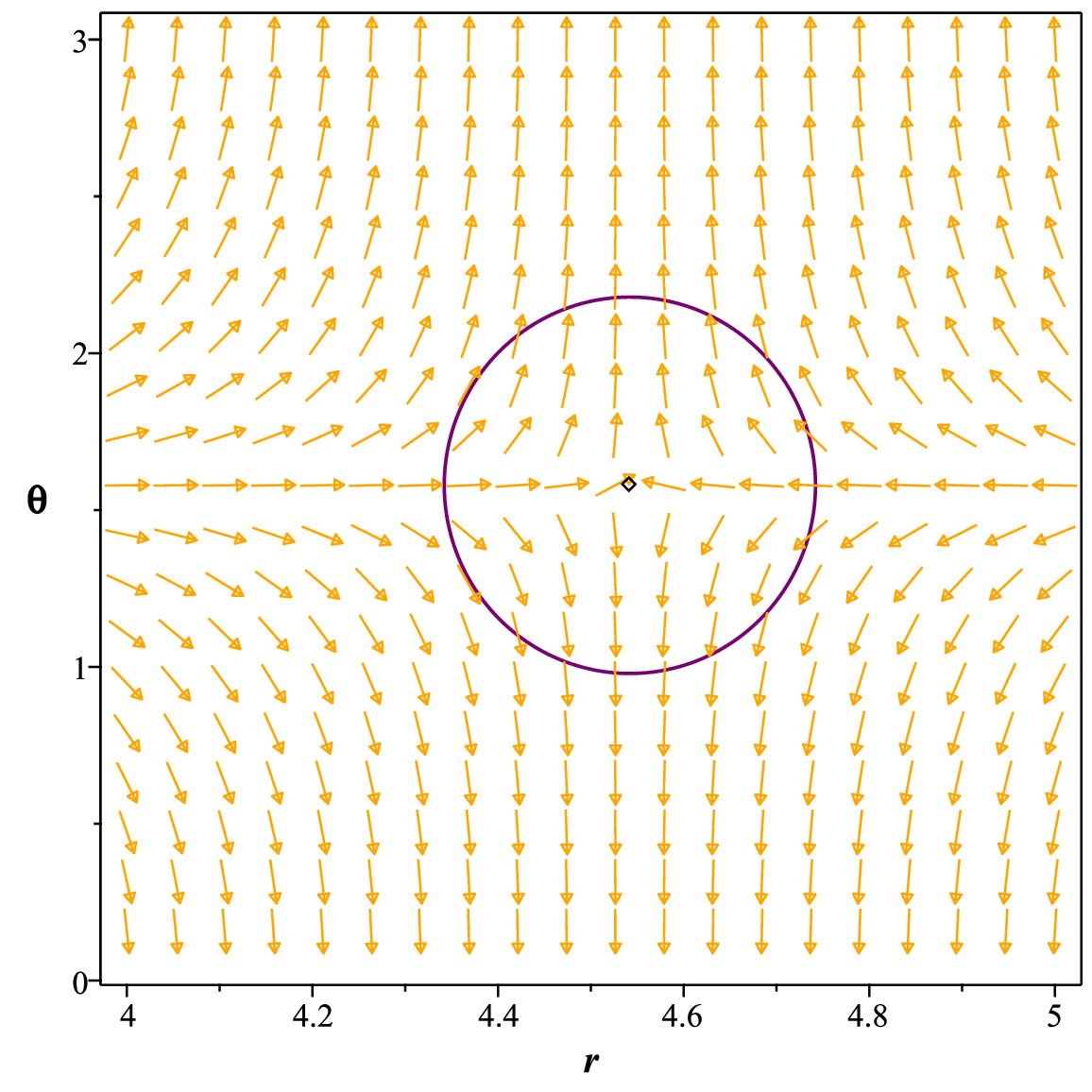}
 \label{5a}}
 \subfigure[]{
 \includegraphics[height=6.5cm,width=8cm]{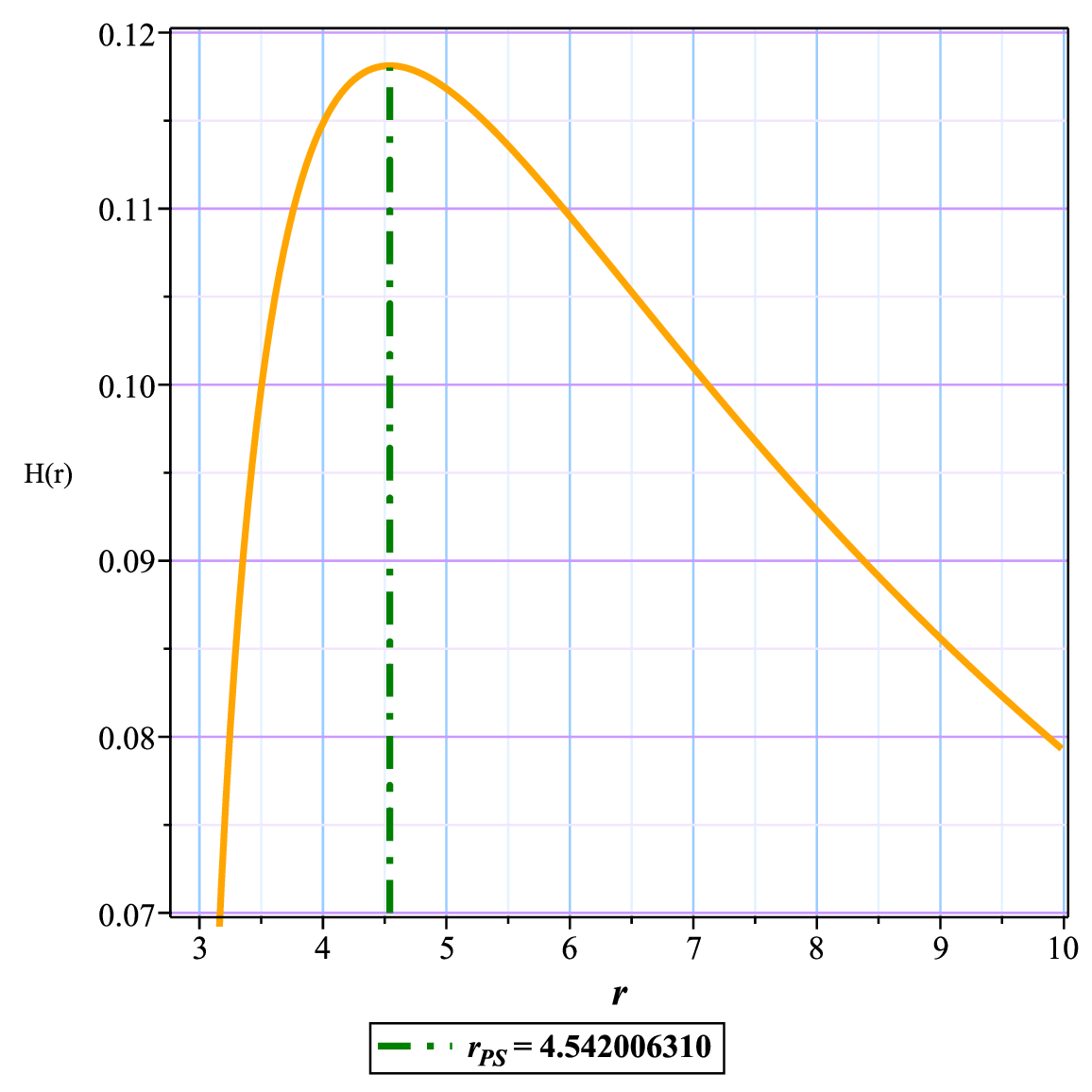}
 \label{5b}}
 
   \caption{\small{Fig (5a):The normal vector field $n$ in the $(r-\theta)$ plane. The photon sphere is located at $ (r,\theta)=(4.542006310,1.57)$ with respect to $( \lambda=0.62, M=1 )$  , (5b): the topological potential H(r) for PFDM black hole model  }}
 \label{2}
\end{center}
\end{figure}
In the first case, considering the chosen value of the parameter $\lambda=0.62 $ and the presence of the event horizon ($r_h=2.97163$) Fig (4), we observe the appearance of a photon sphere outside the event horizon. This photon sphere possesses a topological charge of -1 Fig (5a) and, as evident in Fig (5b), represents an energy maximum. Consequently, in this case, we are confronted with a black hole that contains an unstable photon sphere. 
However, in the second case, our choice of $\lambda=2.15 $ results in a metric function without roots, which is clearly visible in Fig (4). The gravitational structure displays two photon sphere with charges of -1 and +1, indicating a space-time with a TTC of zero, Fig (6a). Energy-wise, this situation is equivalent to the emergence of both a minimum and a maximum, as shown in Fig (6b). Given these conditions, in this scenario, the structure manifests itself in the form of a naked singularity.\\Now we have to answer the question that for what values of the parameter $\lambda$ the space will be in the form of a black hole or naked singularity?
We have shown the answers to these questions in Table 1. Here, it is necessary to note that we have focused on the domain of $\lambda$ and assigned an arbitrary value to M. Obviously, changing these values can change our parameter range.
\begin{center}
\textbf{Case II: TTC = 0 }
\end{center}
\begin{figure}[H]
 \begin{center}
 \subfigure[]{
 \includegraphics[height=6.5cm,width=8cm]{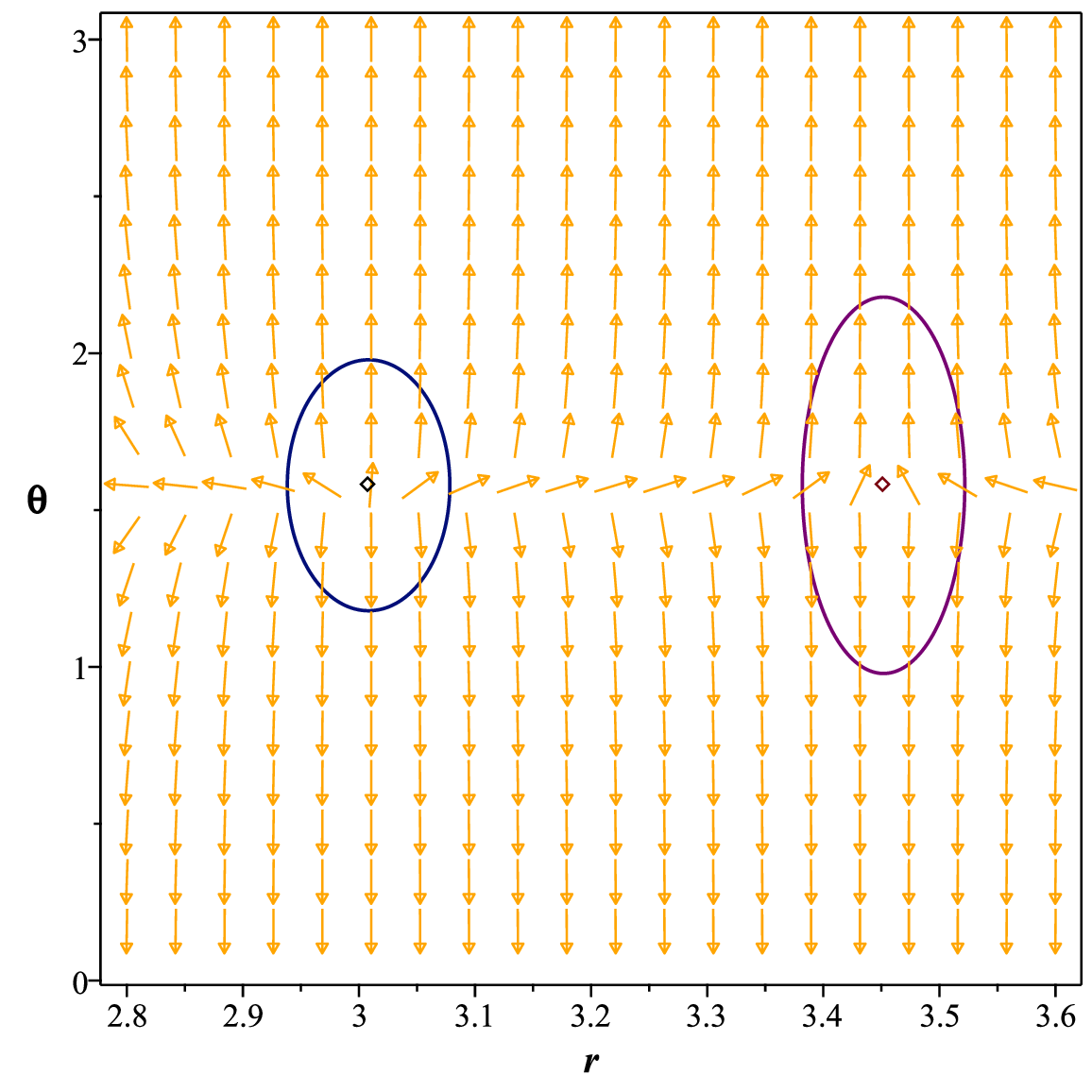}
 \label{5a}}
 \subfigure[]{
 \includegraphics[height=6.5cm,width=8cm]{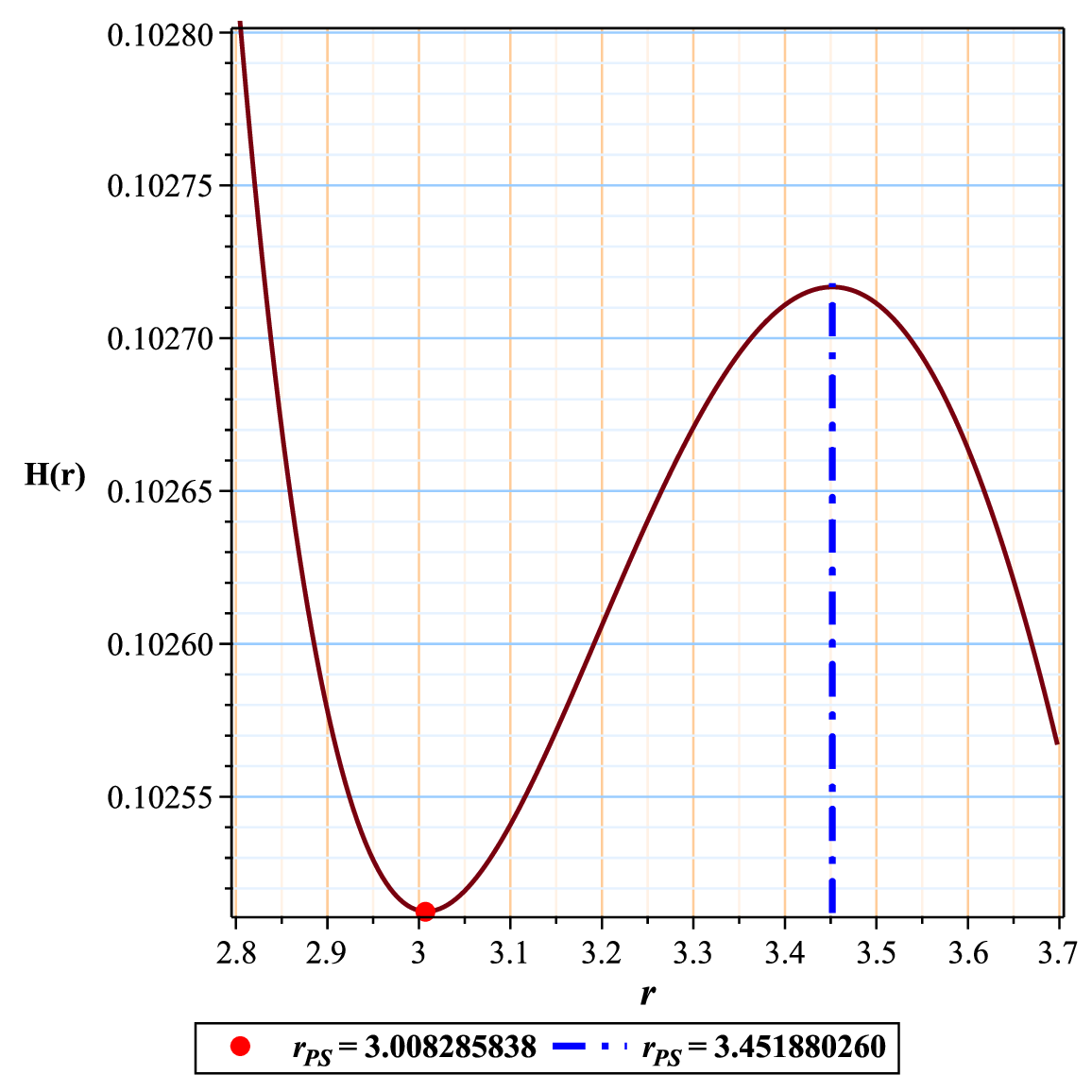}
 \label{5b}}
 
   \caption{\small{Fig (6a):The normal vector field $n$ in the $(r-\theta)$ plane. The photon spheres are located at $ (r,\theta)=(3.008285838,1.57)$ and $ (r,\theta)=(3.451880260,1.57)$  with respect to $( \lambda=2.15, M=1 )$  , (6b): the topological potential H(r) for PFDM black hole model  }}
 \label{2}
\end{center}
\end{figure}

\begin{center}
\begin{table}[H]
  \centering
\begin{tabular}{|p{3cm}|p{4cm}|p{5cm}|p{1.5cm}|p{2cm}|}
  \hline
  \centering{PFDM black holes}  & \centering{Fix parametes} &\centering{Conditions}& *TTC&\ $(R_{PLPS})$\\[3mm]
   \hline
  \centering{*Unauthorized area} & \centering $ M=1 $ & \centering{$\lambda\leq 0 ,\lambda>2.1554784 $} & $nothing$&\ $-$\\[3mm]
   \hline
 \centering{unstable photon sphere} & \centering $M=1$ & \centering{$0<\lambda < 2 $} &\centering $-1$&\ $2.019966202$ \\[3mm]
   \hline
   \centering{naked singularity} & \centering $M=1$ & \centering{$2<\lambda \leq 2.1554784$} & \centering $ 0 $ &\ $-$ \\[3mm]
   \hline
   \end{tabular}
   \caption{*Unauthorized region: is the region where the roots of $\phi$ equations become negative or imaginary in this region\\TTC: *Total Topological Charge}\label{1}
\end{table}
 \end{center}

\subsection{Photon Sphere and charged AdS black hole with perfect fluid dark matter(CPFDM)}
In order to observe the effects of adding the electric field and the AdS radius to the action of the PDFM model on the sphere-topological photon structure, this time we go to the CPFDM model in the AdS form \cite{48}.The metric for such black hole is
\begin{equation}\label{(20)}
f \! \left(r \right)=1-\frac{2 m}{r}+\frac{q^{2}}{r^{2}}+\frac{r^{2}}{l^{2}}+\frac{\ln \! \left(\frac{r}{\lambda}\right) \lambda}{r},
\end{equation}
where m is mass, q is charge, $ \lambda$ is the parameter of intensity of PFDM, and $l$ is the AdS radius length.\\
With respect to(21),(22) and from equations (15) and (16), we will have:
\begin{equation}\label{(2)}
H =\frac{\sqrt{1-\frac{2 m}{r}+\frac{q^{2}}{r^{2}}+\frac{r^{2}}{l^{2}}+\frac{\ln \left(\frac{r}{\lambda}\right) \lambda}{r}}}{\sin \! \left(\theta \right) r},
\end{equation}

\begin{equation}\label{(3)}
\phi^{r}=\frac{\left(-3 \ln \! \left(\frac{r}{\lambda}\right) \lambda  r -2 r^{2}+\left(6 m +\lambda \right) r -4 q^{2}\right) \sqrt{1-\frac{2 m}{r}+\frac{q^{2}}{r^{2}}+\frac{r^{2}}{l^{2}}+\frac{\ln \left(\frac{r}{\lambda}\right) \lambda}{r}}}{2 \sqrt{\frac{\ln \left(\frac{r}{\lambda}\right) \lambda  r \,l^{2}+\left(-2 m r +q^{2}+r^{2}\right) l^{2}+r^{4}}{r^{2} l^{2}}}\, r^{4} \sin \! \left(\theta \right)},
\end{equation}
\begin{equation}\label{(4)}
\phi^{\theta}=-\frac{\sqrt{1-\frac{2 m}{r}+\frac{q^{2}}{r^{2}}+\frac{r^{2}}{l^{2}}+\frac{\ln \left(\frac{r}{\lambda}\right) \lambda}{r}}\, \cos \! \left(\theta \right)}{\sin \! \left(\theta \right)^{2} r^{2}}.
\end{equation}
Here, we have focused on $\lambda$ and assigned arbitrary values to m, $l$, and q. It is clear that changing these values can cause a change in our parameter range.\\
As it is evident from table (2), the forbidden regions are located next to the allowed regions and for this black hole there is practically no region that is equivalent to the naked singularity in terms of total topological charges.
As seen in Fig (7a), the structure shows the behavior of a normal black hole with a total charge of -1 and an unstable photon sphere, which is confirmed with a maximum energy Fig (7b). In comparison with the results from the previous state, it appears that the addition of an electric field and the AdS radius to the action has diminished the impact of the PDFM term. This is because changes in the $\lambda$ parameter, which previously led to the emergence of energy minima beyond the event horizon and drove the model towards a naked singularity, now have no effect.
\begin{center}
\begin{table}[H]
  \centering
\begin{tabular}{|p{3cm}|p{4cm}|p{4cm}|p{4cm}|}
  \hline
  \centering{CPFDM black holes}  & \centering{Fix parametes} &\centering{Conditions}&*TTC\\[3mm]
   \hline
  \centering{*Unauthorized area} & $q=0.1,m=1,l=1$ & \centering{$\lambda<0$} &$nothing$\\[3mm]
   \hline
  \centering{unstable photon sphere} & $q=0.1,m=1,l=1$ & \centering{$0<\lambda$} &$-1$\\[3mm]
   \hline
      \end{tabular}
   \caption{*Unauthorized region: is the region where the roots of $\phi$ equations become negative or imaginary in this region.\\ TTC: *Total Topological Charge\\}\label{1}
\end{table}
 \end{center}
\begin{figure}[H]
 \begin{center}
 \subfigure[]{
 \includegraphics[height=6.5cm,width=8cm]{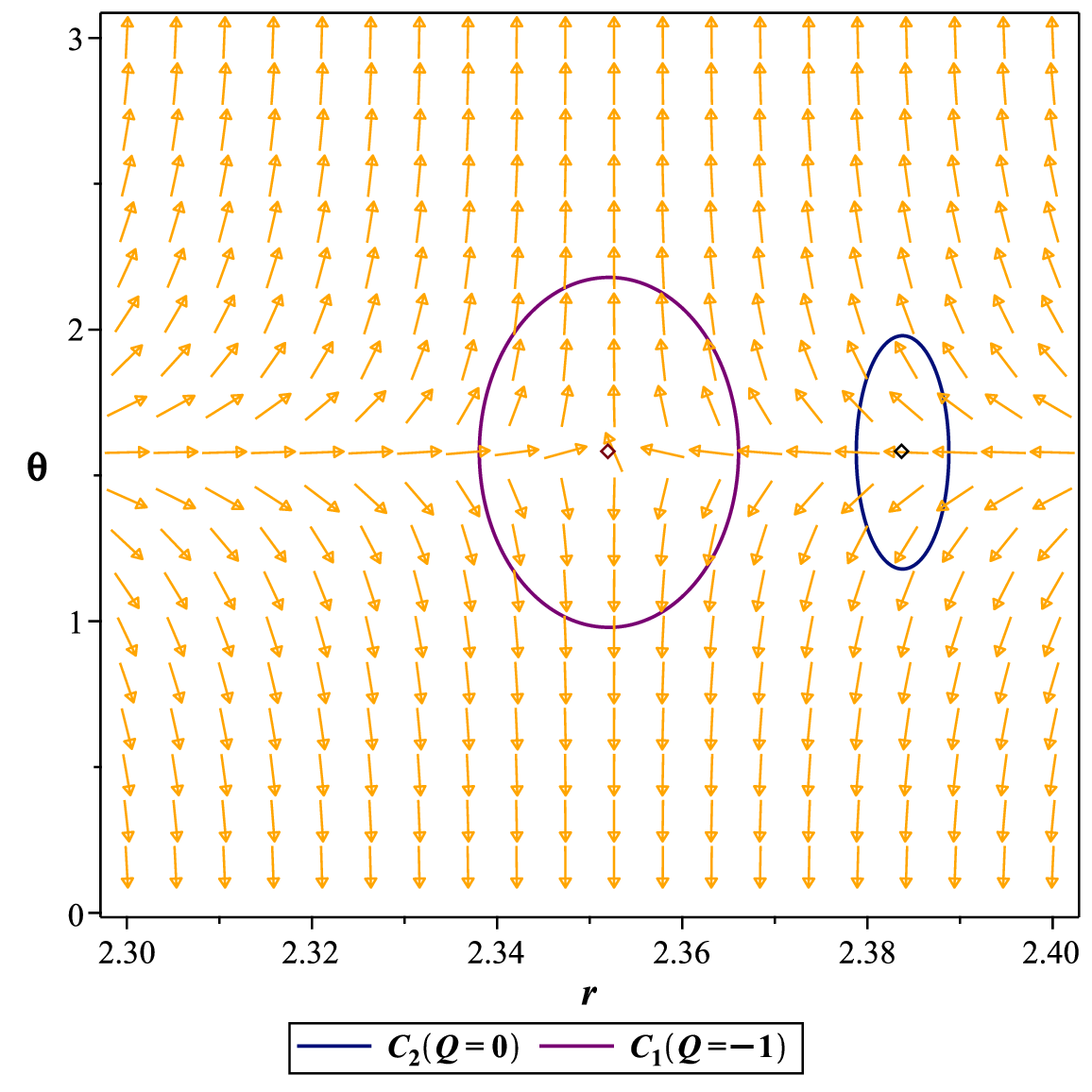}
 \label{7a}}
 \subfigure[]{
 \includegraphics[height=6.5cm,width=8cm]{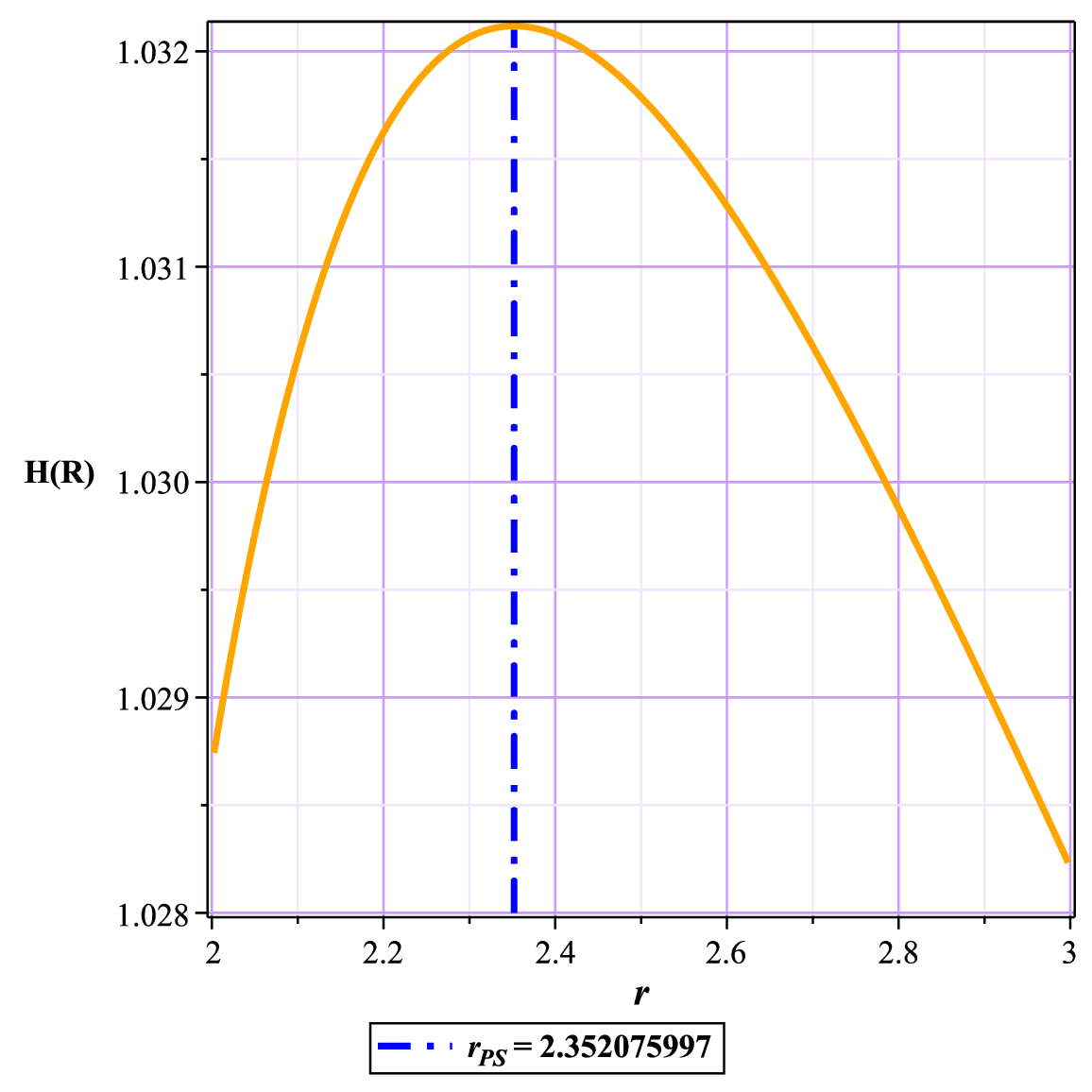}
 \label{7b}}
  \caption{\small{The normal vector field $n$ in the $(r-\theta)$ plane. The photon spheres are located at $(r,\theta)=(2.352075997,1.57)$ for Fig (7a) with respect to $(q=0.1,m=1,\lambda=0.2,l=1)$, (7b): the topological potential H(r) for CPFDM black hole model }}
 \label{7}
\end{center}
 \end{figure}
\subsection{Photon Sphere and Euler-Heisenberg black hole}
The Euler-Heisenberg black hole is a type of black hole that has a nonlinear electromagnetic field due to quantum effects.
In fact, Euler and Heisenberg developed an effective Lagrangian for quantum electrodynamics (QED) that described the interaction of photons with each other in the presence of a strong electromagnetic field\cite{50}
Casimir and Polder applied Euler-Heisenberg Lagrangian to the problem of a charged black hole in general relativity. They found that QED corrections  affect the electromagnetic field,  thermodynamics and stability of black hole . They named this black hole Euler-Heisenberg black hole and showed that it has interesting properties.\\
• The QED parameter, which measures the strength of the nonlinear effects, can be positive or negative. A positive QED parameter reduces the electric field near the horizon and increases the mass of the black hole, while a negative QED parameter enhances the electric field and decreases the mass\cite{51}.\\
• The QED parameter also influences the existence and location of the inner and outer horizons of the black hole. For a positive QED parameter, there is always a single horizon, while for a negative QED parameter, there can be two horizons, one horizon, or no horizon at all, depending on the values of the mass and charge\cite{51}.\\
• The QED parameter also modifies the stability of the black hole against perturbations. For a positive QED parameter, the black hole is stable for any mass and charge, while for a negative QED parameter, there is a range of mass and charge where the black hole is unstable and can decay into radiation or smaller black holes\cite{52}.\\
Euler Heisenberg's spherical four-dimensional black hole has a metric function in the following form\cite{53}:
\begin{figure}[H]
 \begin{center}
 \subfigure[]{
 \includegraphics[height=6.5cm,width=8cm]{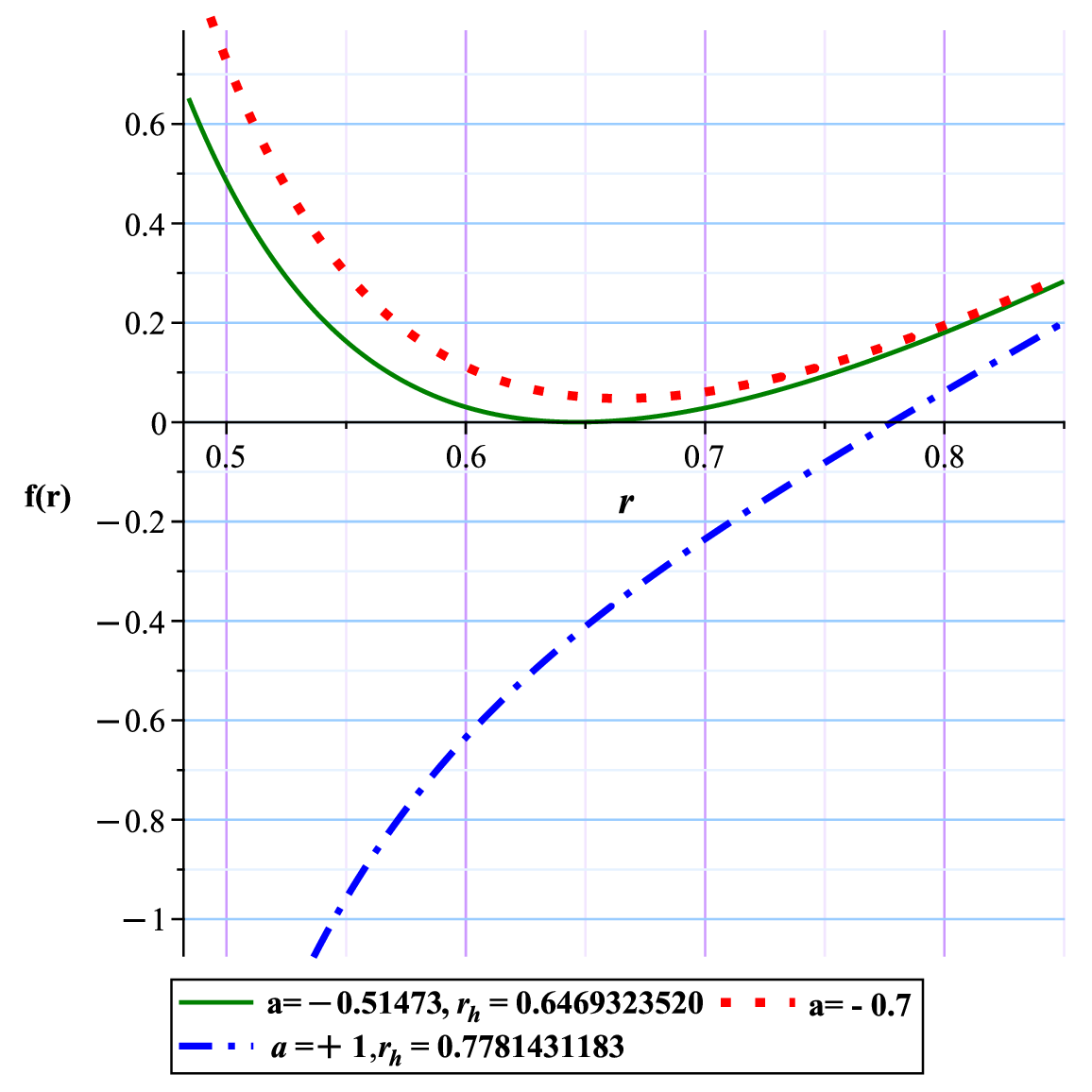}
 \label{8a}}
 
   \caption{\small{Metric function with different $a$ for Euler-Heisenberg black hole model  }}
 \label{8}
\end{center}
\end{figure}

\begin{equation}\label{20}
\begin{split}
f(r)=1-\frac{2m}{r}+\frac{q^2}{r^2}+\frac{r^2}{l^2}-\frac{a q^4}{20 r^6},
\end{split}
\end{equation}
where parameter m is the ADM mass, q is the electric charge and "a" represents the strength of the QED correction.
With respect to(21),(22) and from equations (15) and (16), we will have,
\begin{equation}\label{23}
\begin{split}
H =\frac{\sqrt{100-\frac{200 m}{r}+\frac{100 q^{2}}{r^{2}}+\frac{100 r^{2}}{l^{2}}-\frac{5 q^{4} a}{r^{6}}}}{10 \sin \! \left(\theta \right) r},
\end{split}
\end{equation}
\begin{equation}\label{24}
\begin{split}
\phi^{r}=\frac{\left( 30.0 r^{5}- 12.8 r^{4}+ 0.8192 a - 10.0 r^{6}\right) \sqrt{100-\frac{200 m}{r}+\frac{100 q^{2}}{r^{2}}+\frac{100 r^{2}}{l^{2}}-\frac{5 q^{4} a}{r^{6}}}}{10 r^{8} \sin \! \left(\theta \right) \sqrt{\frac{\left( 100.0 r^{6}- 200.0 r^{5}+ 64.0 r^{4}- 2.048 a \right) l^{2}+ 100.0 r^{8}}{r^{6} l^{2}}}},
\end{split}
\end{equation}
\begin{equation}\label{(25)}
\phi^{\theta}=-\frac{\sqrt{\frac{\left(-200 r^{5} m +100 r^{4} q^{2}+100 r^{6}-5 q^{4} a \right) l^{2}+100 r^{8}}{l^{2} r^{6}}}\, \cos \! \left(\theta \right)}{10 \sin \! \left(\theta \right)^{2} r^{2}}.
\end{equation}
\begin{center}
\textbf{Case I: TTC =-1}
\end{center}
\begin{figure}[H]
 \begin{center}
 \subfigure[]{
 \includegraphics[height=6.5cm,width=8cm]{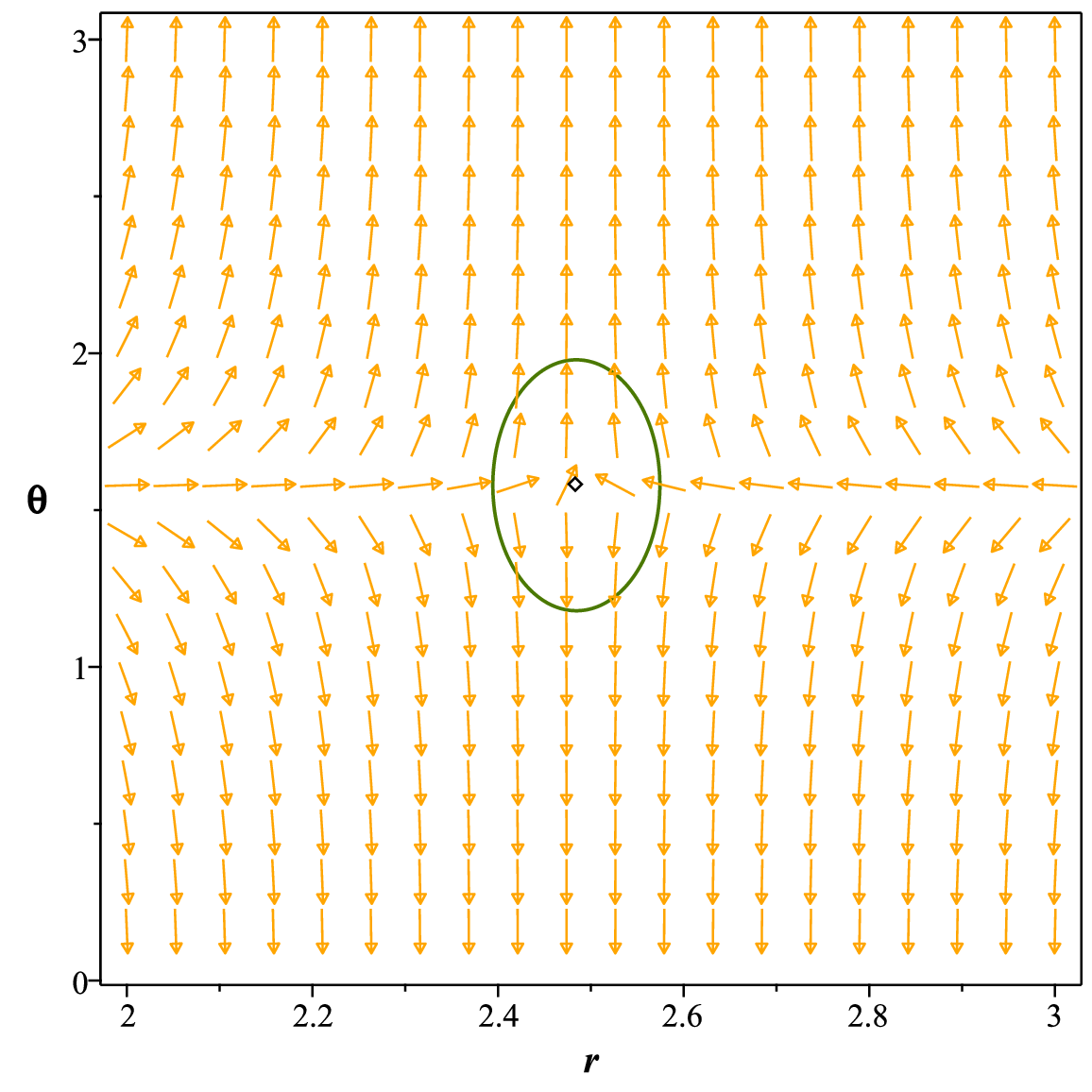}
 \label{9a}}
 \subfigure[]{
 \includegraphics[height=6.5cm,width=8cm]{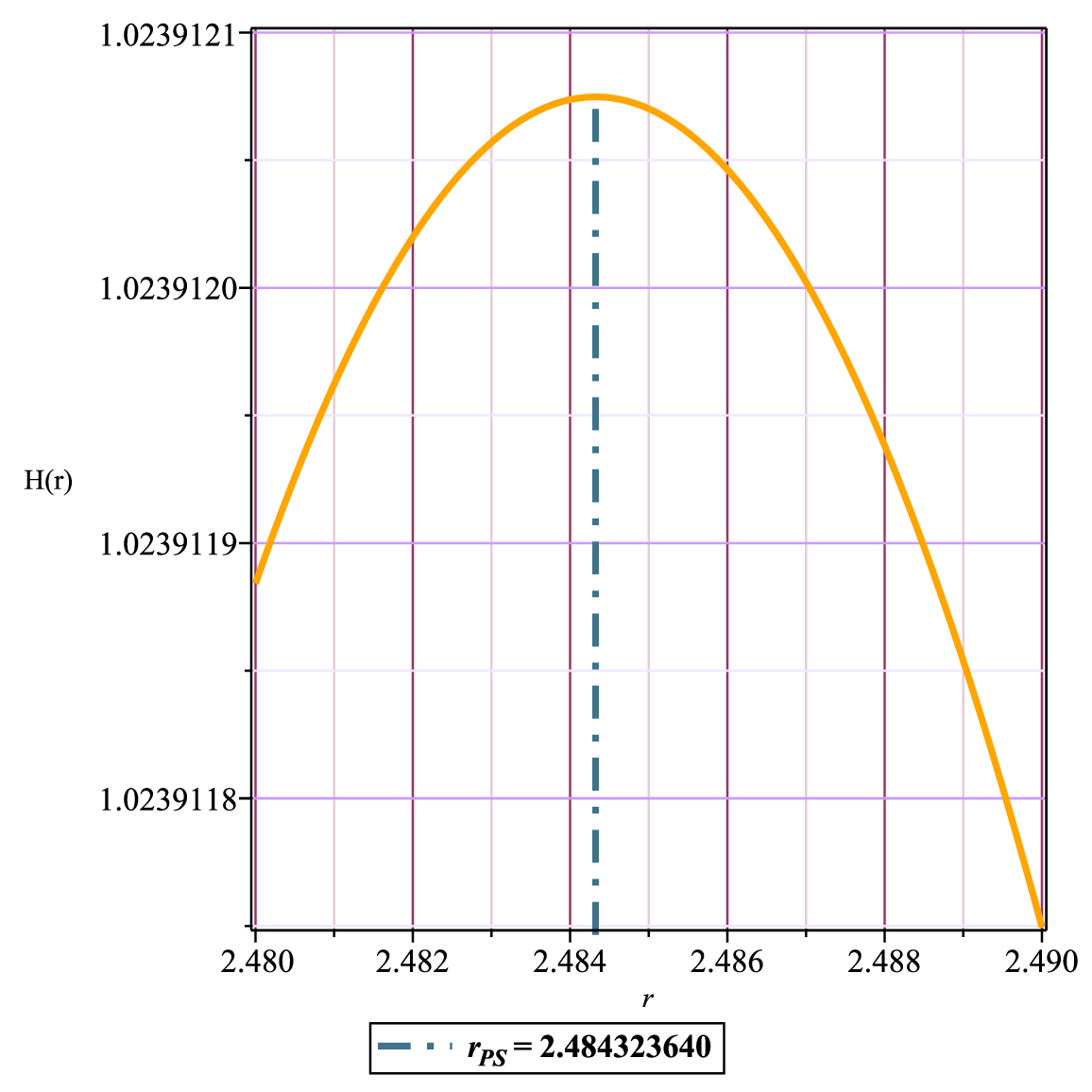}
 \label{9b}}
  \caption{\small{The normal vector field $n$ in the $(r-\theta)$ plane. The photon spheres are located at $ (r,\theta)=(2.484328812,1.57)$ for Fig (9a) with respect to $(q=0.8,m=1,a=-0.51473,l=1)$, Fig (9b) the topological potential H(r) for Euler-Heisenberg black hole model}}
 \label{9}
\end{center}
 \end{figure}
 
 \begin{center}
\textbf{Case II: TTC = 0 }
\end{center}
 
\begin{figure}[H]
 \begin{center}
 \subfigure[]{
 \includegraphics[height=6.5cm,width=8cm]{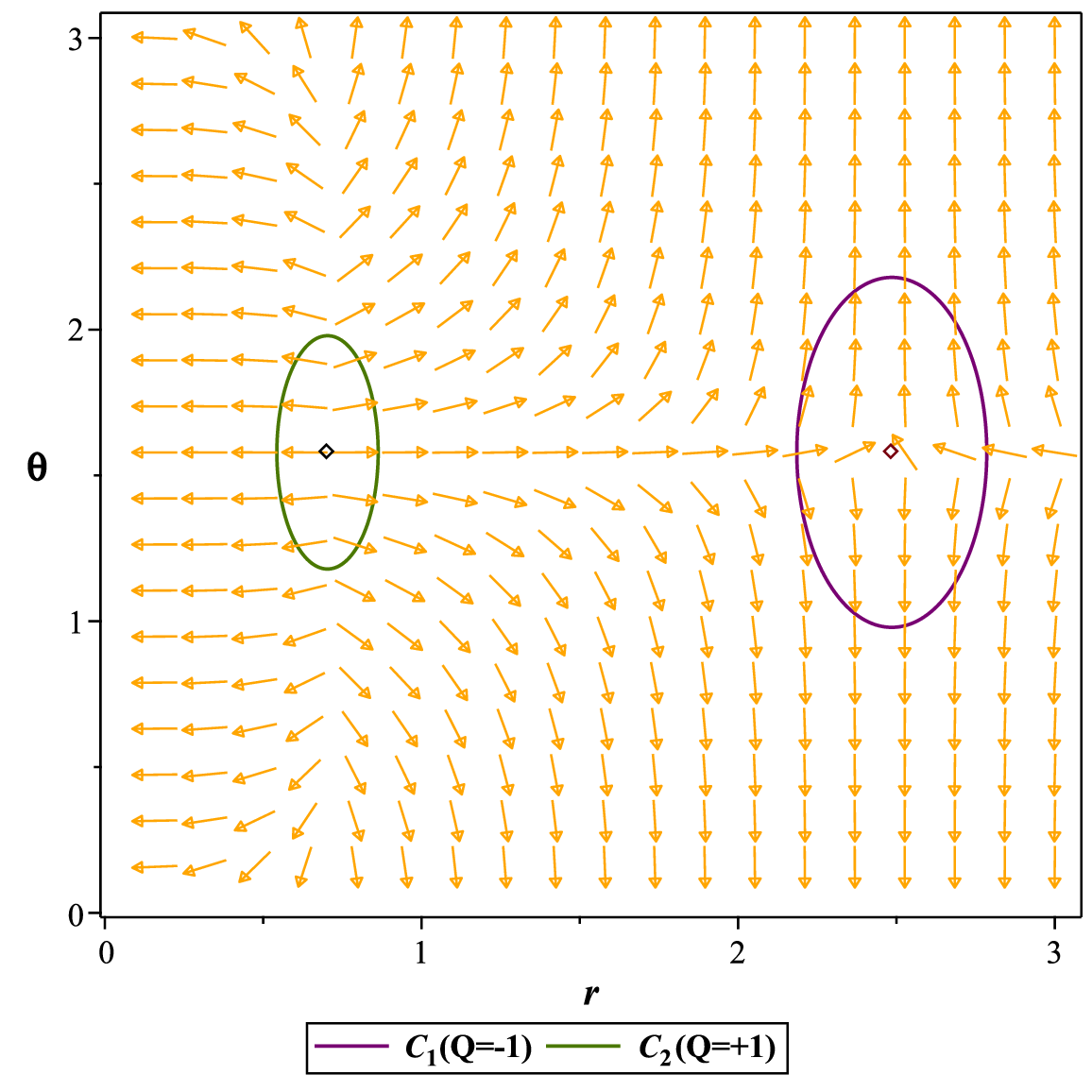}
 \label{10a}}
 \subfigure[]{
 \includegraphics[height=6.5cm,width=8cm]{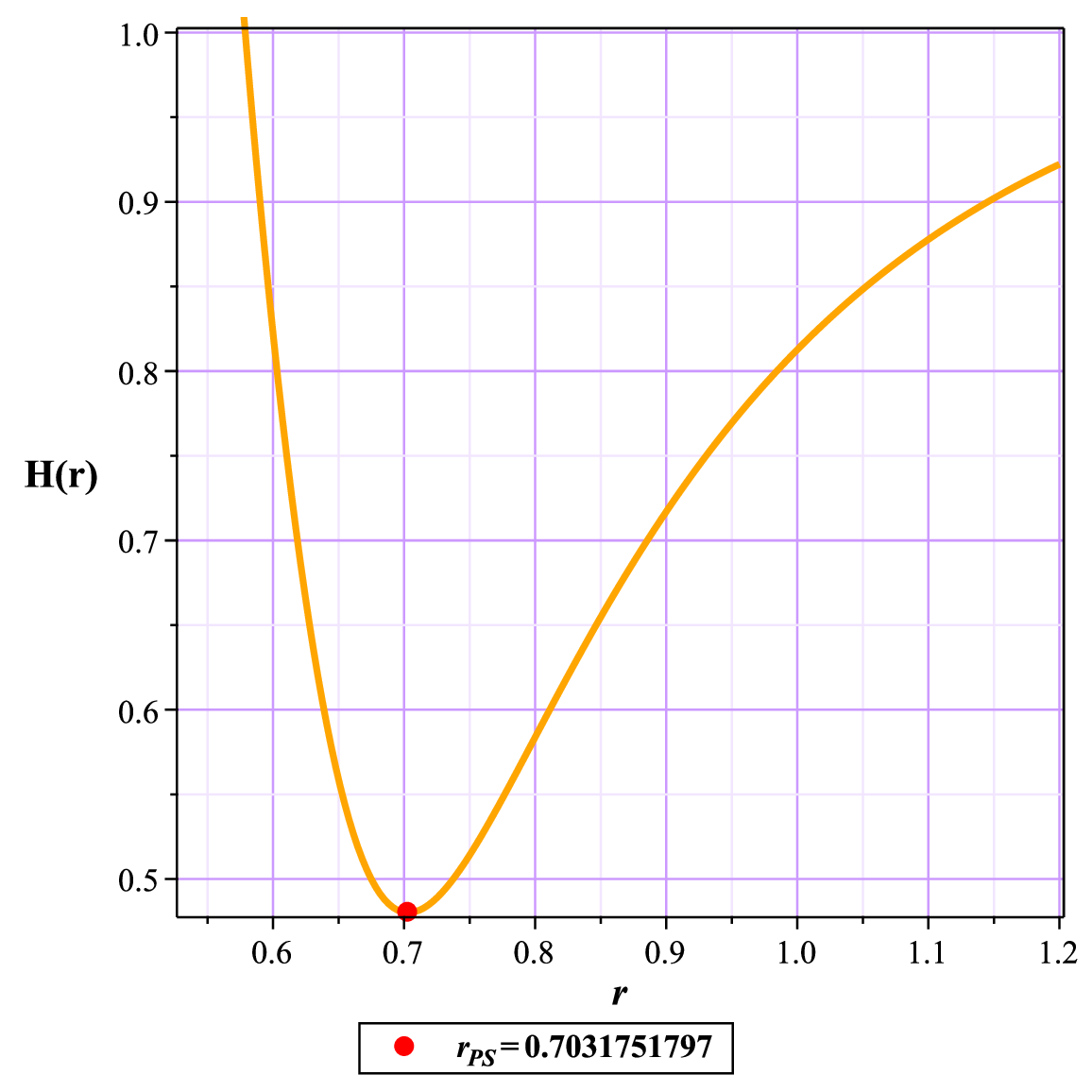}
 \label{10b}}
 \subfigure[]{
 \includegraphics[height=6.5cm,width=8cm]{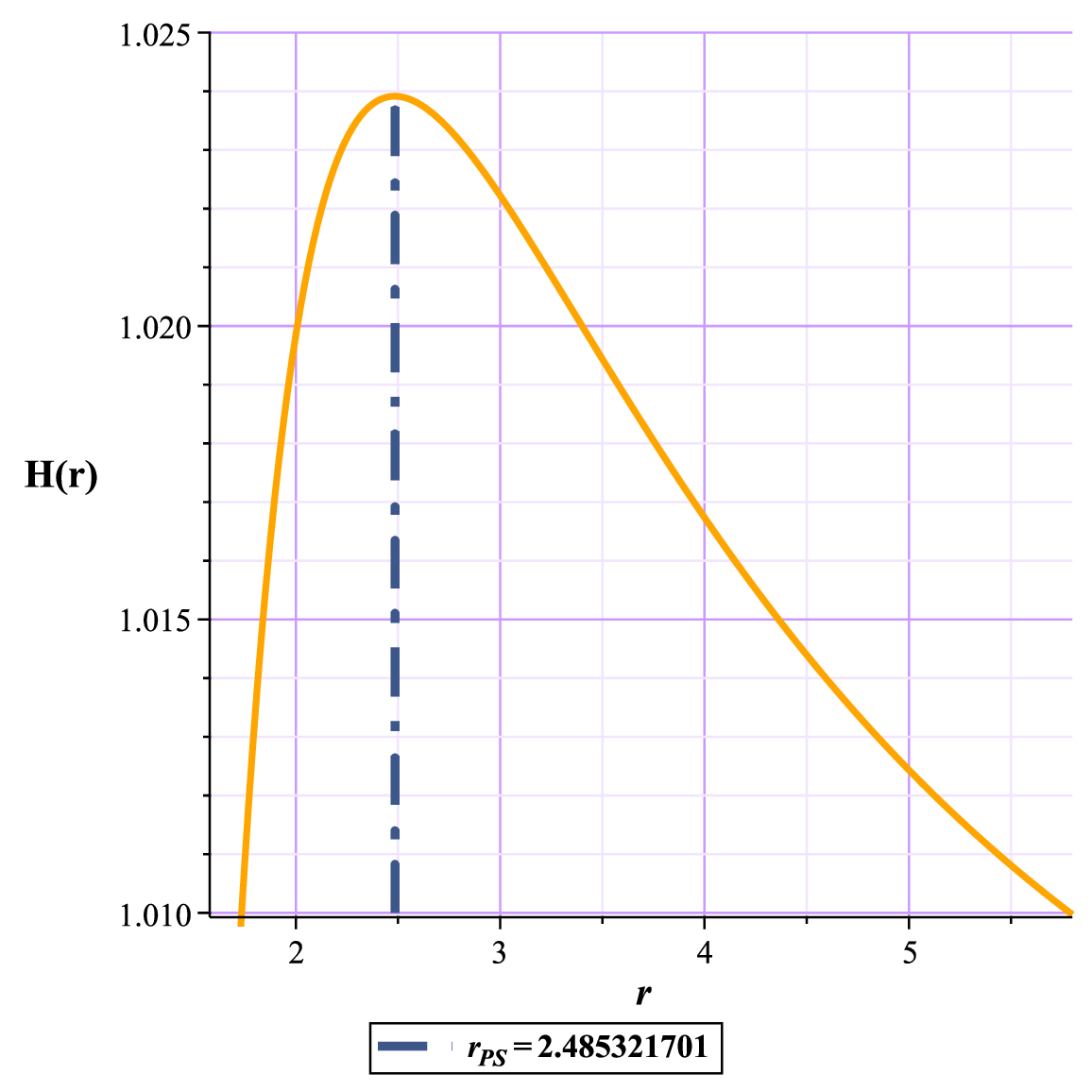}
 \label{10c}}
  \caption{\small{The normal vector field $n$ in the $(r-\theta)$ plane. The photon spheres are located at  $(r,\theta)=(0.7031751797,1.57)$, $(r,\theta)=(2.483792443,1.57)$ for Fig (10a) with respect to $(q=0.8,m=1,a=-1,l=1)$, Figs (10b),(10c) the topological potential H(r) for Euler-Heisenberg black hole model }}
 \label{10}
\end{center}
 \end{figure} 
 
We have focused on parameter $a$, for this purpose we have considered an arbitrary value for mass and charge $(m= 1 , q= 0.8)$.
As we can see in Figs (8),(9),(10), our chosen range for $a$ can lead to different behaviors for the photon sphere.
For example, Fig (9a), with a topological charge of -1, indicates a normal black hole behavior, it means that  we have an unstable photon sphere.
Conversely, in Fig (10b), the appearance of a minimum in the space without an event horizon and the zero TTC Fig (10a) all imply the existence of a naked singularity.
Based on this, the following intervals can be used to have different modes (table3).

\begin{center}
\begin{table}[H]
  \centering
\begin{tabular}{|p{3cm}|p{4cm}|p{5cm}|p{1.5cm}|p{2cm}|}
  \hline
  \centering{Euler-Heisenberg-AdS black holes}  & \centering{Fix parametes} &\centering{Conditions}& *TTC&\ $(R_{PLPS})$\\[3mm]
   \hline
  \centering{*Unauthorized area} & $q=0.8,m=1,l=1$ & \centering{$a <- 144.87$} & $nothing$&\ $-$\\[3mm]
   \hline
  \centering{naked singularity} & $q=0.8,m=1,l=1$ & \centering{$- 144.87<a <-0.51473$} & $0$&\ $-$ \\[3mm]
   \hline
   \centering{unstable photon sphere} & $q=0.8,m=1,l=1$ & \centering{$a\geq-0.51473$} & $-1$&\ $2.484323651$ \\[3mm]
   \hline
   \end{tabular}
   \caption{*Unauthorized region: is the region where the roots of $\phi$ equations become negative or imaginary in this region\\TTC: *Total Topological Charge}\label{1}
\end{table}
 \end{center}
 \subsection{Photon Sphere and Euler-Heisenberg black hole surrounded by PFDM}
In order to observe the effects of adding the PFDM to the action of the Euler-Heisenberg model on the sphere-topological photon structure, this time we go to the Euler-Heisenberg black hole surrounded by PFDM model \cite{54}. The metric for such black hole is:
\begin{equation}\label{20}
\begin{split}
f =1-\frac{2 m}{r}+\frac{q^{2}}{r^{2}}-\frac{a \,q^{4}}{20 r^{6}}+\frac{\lambda  \ln \! \left(\frac{r}{\lambda}\right)}{r},
\end{split}
\end{equation}
\begin{figure}[H]
 \begin{center}
 \subfigure[]{
 \includegraphics[height=6.5cm,width=8cm]{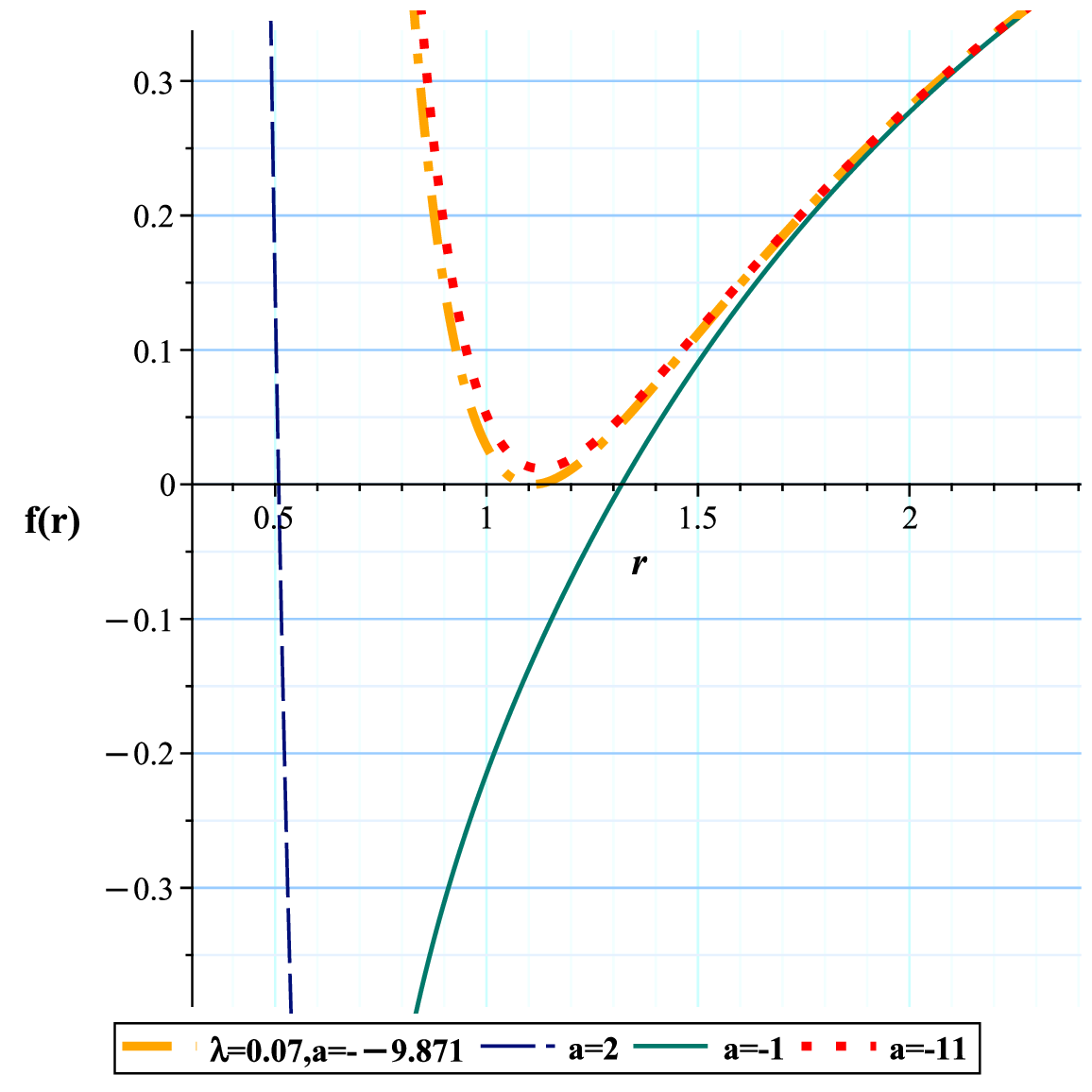}
 \label{8a}}
 
   \caption{\small{Metric function with different $a$ for Euler-Heisenberg black hole surrounded by PFDM }}
 \label{8}
\end{center}
\end{figure}
where parameter m is the ADM mass, q is the electric charge ,$ \lambda$ is the parameter of intensity of PFDM and "a" represents the strength of the QED correction.
With respect to(21),(22) and from equations (15) and (16), we will have,
\begin{equation}\label{20}
\begin{split}
H =\frac{\sqrt{100-\frac{200 m}{r}+\frac{100 q^{2}}{r^{2}}-\frac{5 a \,q^{4}}{r^{6}}+\frac{100 \lambda  \ln \left(\frac{r}{\lambda}\right)}{r}}}{10 r \sin \! \left(\theta \right)},
\end{split}
\end{equation}

\begin{equation}\label{20}
\begin{split}
\phi^{r}=-\frac{\left(15 \lambda  \ln \! \left(\frac{r}{\lambda}\right) r^{5}-5 \lambda  r^{5}-30 m \,r^{5}+20 q^{2} r^{4}+10 r^{6}-2 a \,q^{4}\right) \csc \! \left(\theta \right)}{10 r^{8}},
\end{split}
\end{equation}
\begin{equation}\label{20}
\begin{split}
\phi^{\theta}=-\frac{\sqrt{100-\frac{200 m}{r}+\frac{100 q^{2}}{r^{2}}-\frac{5 a \,q^{4}}{r^{6}}+\frac{100 \lambda  \ln \left(\frac{r}{\lambda}\right)}{r}}\, \cos \! \left(\theta \right)}{10 r^{2} \sin \! \left(\theta \right)^{2}}.
\end{split}
\end{equation}
 \begin{center}
\textbf{Case I: TTC =-1}
\end{center}
\begin{figure}[H]
 \begin{center}
 \subfigure[]{
 \includegraphics[height=6cm,width=8cm]{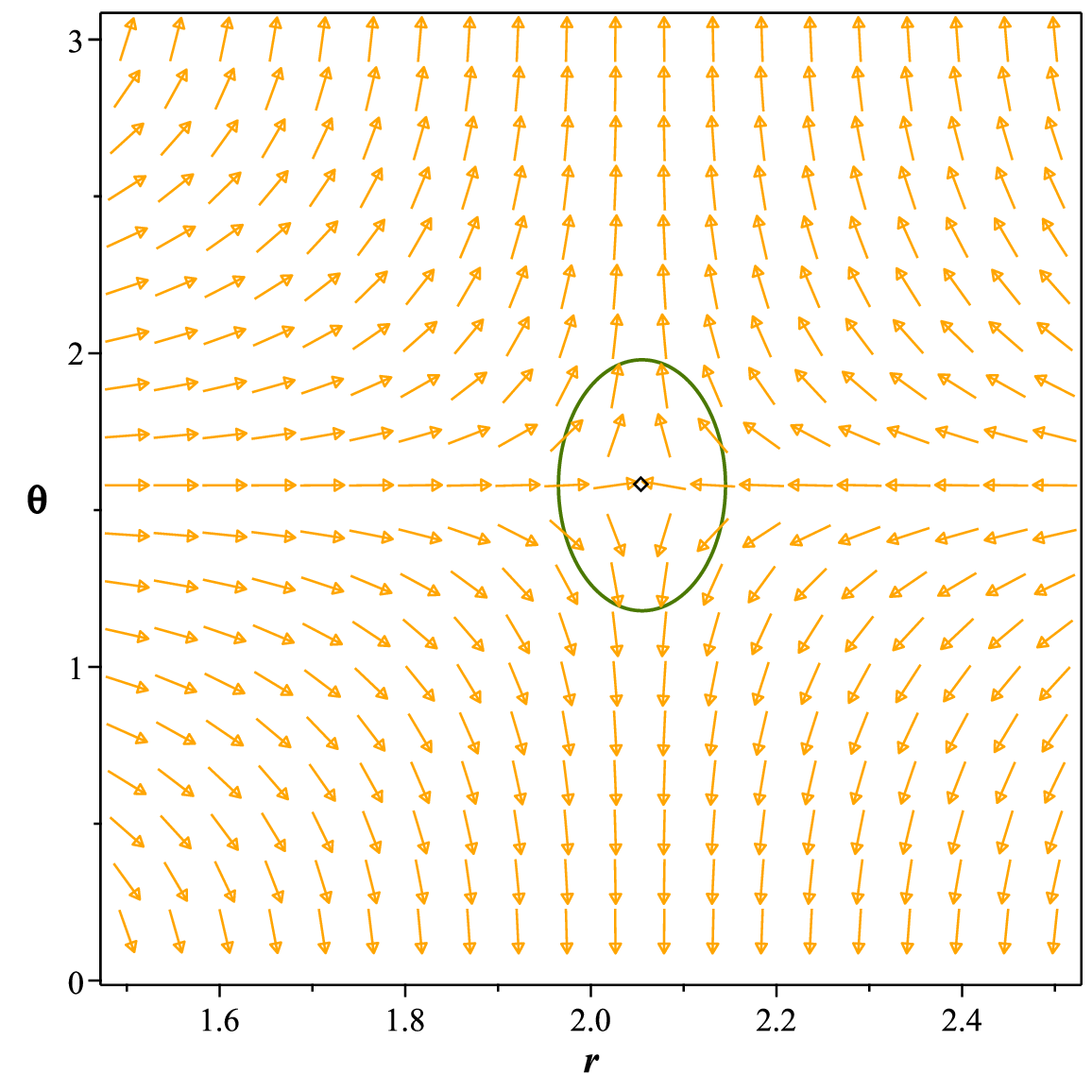}
 \label{9a}}
 \subfigure[]{
 \includegraphics[height=6cm,width=8cm]{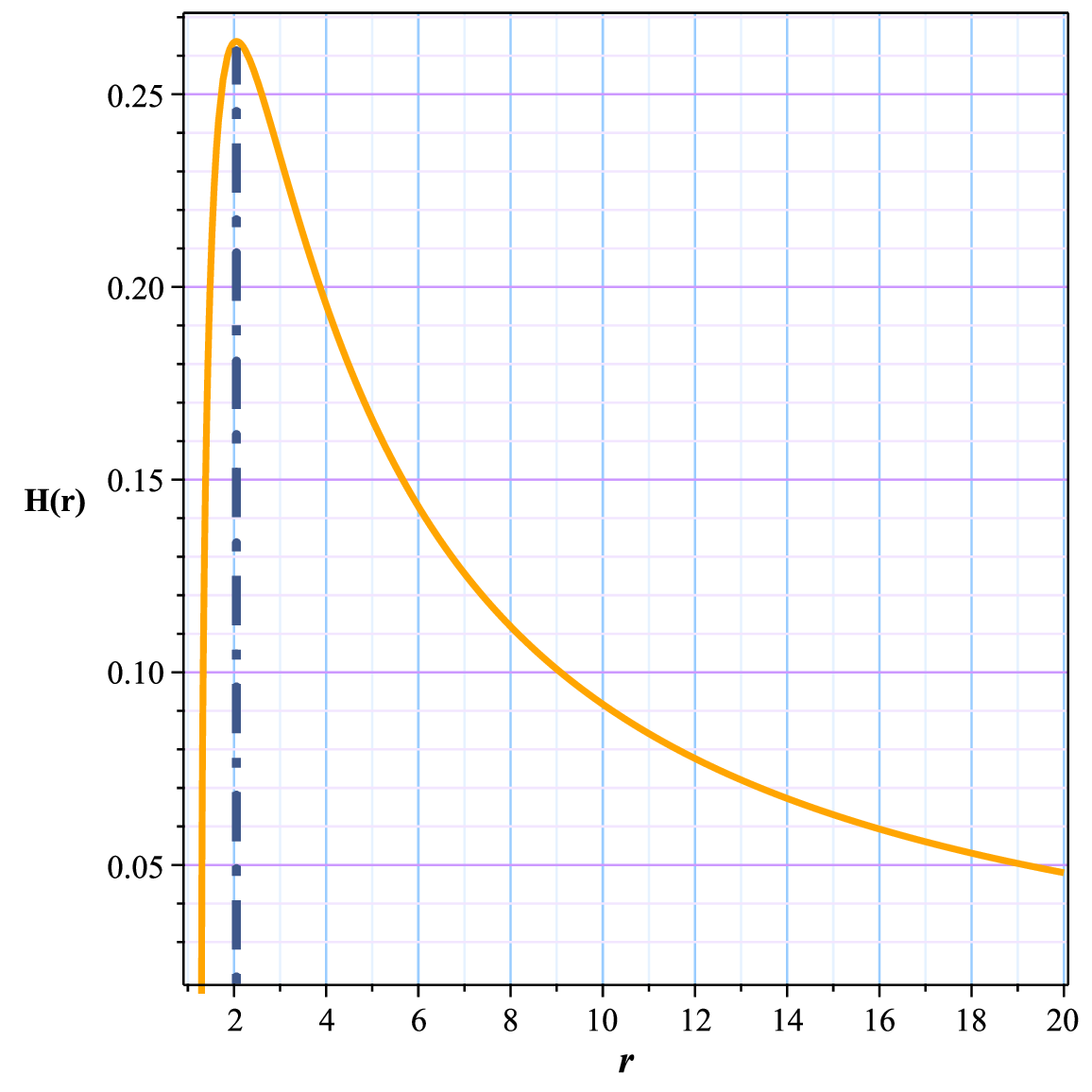}
 \label{9b}}
  \caption{\small{The normal vector field $n$ in the $(r-\theta)$ plane. The photon spheres are located at $ (r,\theta)=(2.055057,1.57)$ for Fig (12a) with respect to $(q=0.8,m=1,a=-0.51473,l=1,\lambda=0.07)$, Fig (12b) the topological potential H(r) for Euler-Heisenberg black hole surrounded by PFDM model}}
 \label{9}
\end{center}
 \end{figure}
Regardless of the impact of the AdS radius, which we have admittedly examined to a lesser extent in this article, it is evident from the comparison of the results in Tables 4 and 5 with Table 3 that the addition of PFDM to the model has caused a significant increase in the range affected by black hole behavior.  
Given the negative value of the parameter (a), which signifies a greater influence of the electric field, it appears that when dark matter is absent from the model (as shown in Table 3), 
even very small values of (a) rapidly expand the domain of gravitational dominance. Consequently, a gravitational minimum emerges swiftly beyond the event horizon, leading to the model's becoming a naked singularity. As observed(in table 3), the domain of dominance of this naked singularity is significantly extensive due to the increase in the parameter within a considerably large negative range. However, in the presence of dark matter (as indicated in Tables 4 and 5), this structure seems to act as a reverse agent, preventing the rapid growth and expansion of the model's gravitational domain. Thus, as evident, we maintain the conditions for the existence of a black hole with an unstable photon sphere over a substantial range of negative (a) values, 
In other words, it appears that the inclusion of dark matter has somewhat prevented the formation of gravitational minima over a larger range, which is an intriguing effect.
\begin{center}
\textbf{Case II: TTC = 0 }
\end{center}
\begin{figure}[H]
 \begin{center}
 \subfigure[]{
 \includegraphics[height=6cm,width=8cm]{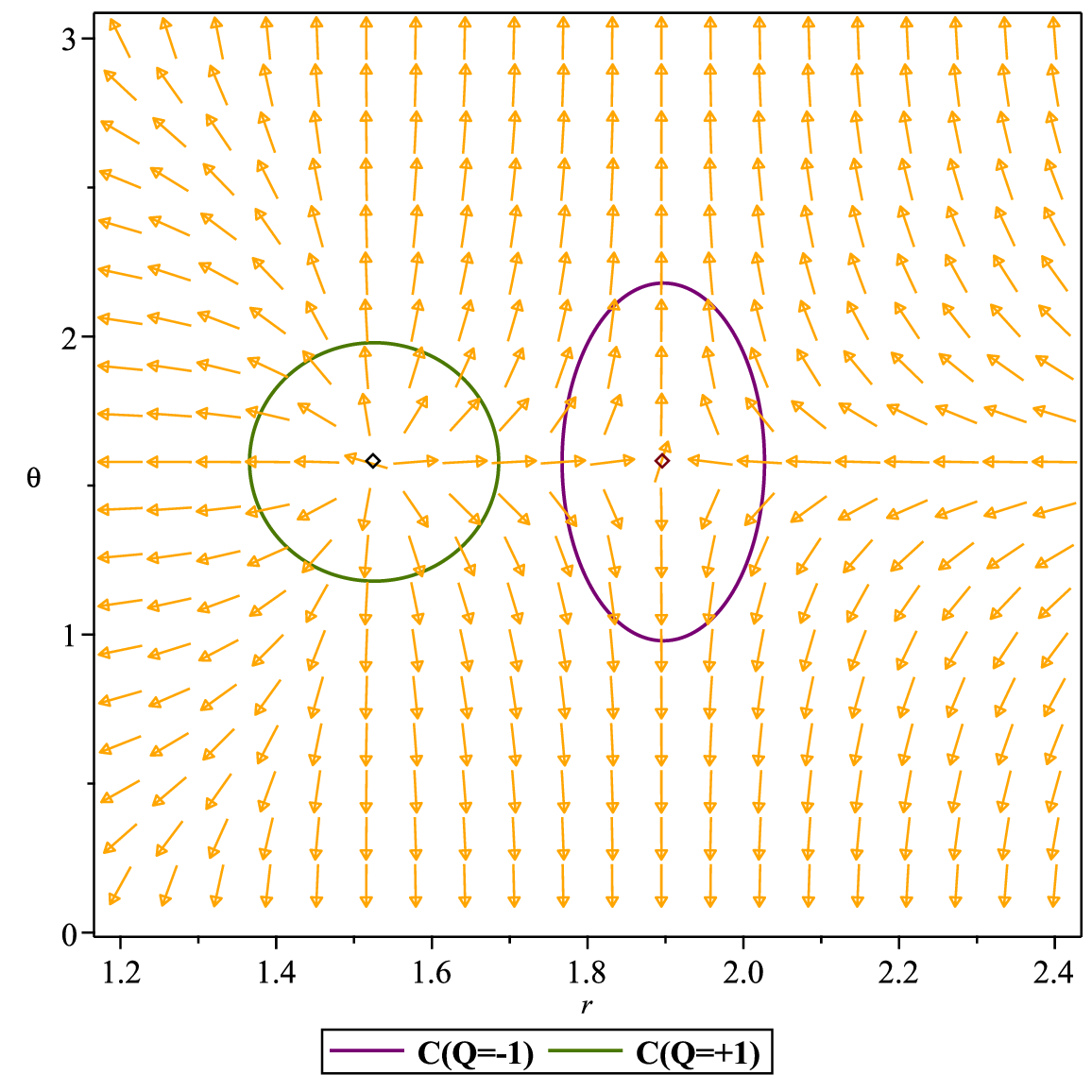}
 \label{10a}}
 \subfigure[]{
 \includegraphics[height=6cm,width=8cm]{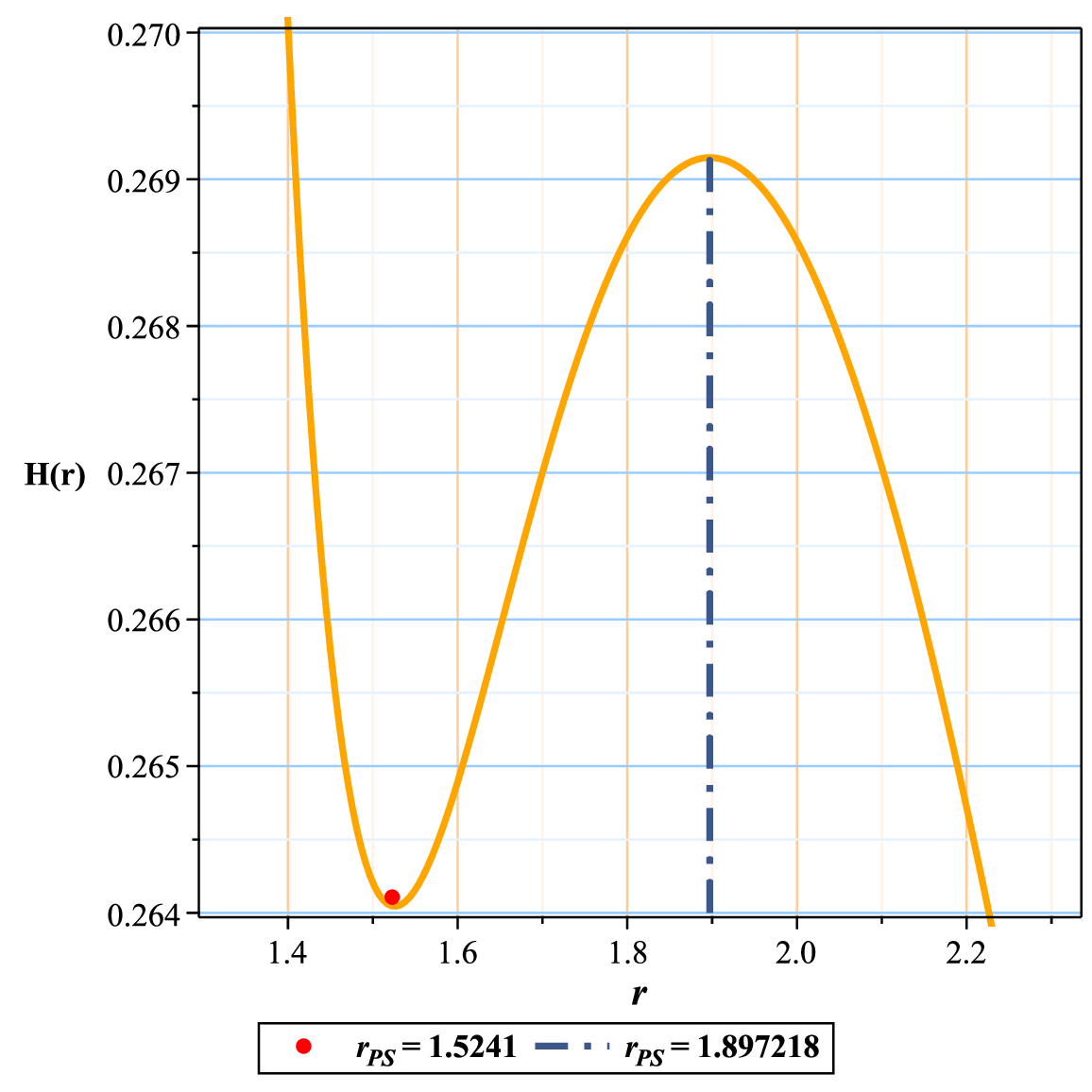}
 \label{10b}}
  \caption{\small{The normal vector field $n$ in the $(r-\theta)$ plane. The photon spheres are located at  $(r,\theta)=(1.5241,1.57)$, $(r,\theta)=(1.897218,1.57)$ for Fig (13a) with respect to $(q=0.8,m=1,a=-1,l=1,\lambda=0.07)$, Fig (13b) the topological potential H(r) for Euler-Heisenberg black hole surrounded by PFDM model }}
 \label{10}
\end{center}
 \end{figure}

 \begin{center}
\begin{table}[H]
  \centering
\begin{tabular}{|p{3cm}|p{4cm}|p{5cm}|p{1.5cm}|p{2cm}|}
  \hline
  \centering{Euler-Heisenberg PFDM black holes}  & \centering{Fix parametes} &\centering{Conditions}& *TTC&\ $(R_{PLPS})$\\[3mm]
   \hline
  \centering{*Unauthorized area} & $q=0.8,m=1,\lambda=0.07$ & \centering{$a <- 40$} & $nothing$&\ $-$\\[3mm]
   \hline
  \centering{naked singularity} & $q=0.8,m=1,\lambda=0.07$ & \centering{$- 39\leq a <-9.872$} &\centering $0$&\ $-$ \\[3mm]
   \hline
   \centering{unstable photon sphere} & $q=0.8,m=1,\lambda=0.07$ & \centering{$a\geq-9.871$} & \centering $-1$&\ $2.026241903$ \\[3mm]
   \hline
   \end{tabular}
   \caption{*Unauthorized region: is the region where the roots of $\phi$ equations become negative or imaginary in this region\\TTC: *Total Topological Charge}\label{1}
\end{table}
 \end{center}

 \begin{center}
\begin{table}[H]
  \centering
\begin{tabular}{|p{3cm}|p{4cm}|p{5cm}|p{1.5cm}|p{2cm}|}
  \hline
  \centering{Euler-Heisenberg PFDM black holes}  & \centering{Fix parametes} &\centering{Conditions}& *TTC&\ $(R_{PLPS})$\\[3mm]
   \hline
  \centering{*Unauthorized area} & $q=0.8,m=1,\lambda=0.6$ & \centering{$a <- 14.9$} & $nothing$&\ $-$\\[3mm]
   \hline
  \centering{naked singularity} & $q=0.8,m=1,\lambda=0.6$ & \centering{$- 14.9<a <-4.21$} &\centering $0$&\ $-$ \\[3mm]
   \hline
   \centering{unstable photon sphere} & $q=0.8,m=1,\lambda=0.6$ & \centering{$a\geq-4.2$} &\centering $-1$&\ $1.583699320$ \\[3mm]
   \hline
   \end{tabular}
   \caption{*Unauthorized region: is the region where the roots of $\phi$ equations become negative or imaginary in this region\\TTC: *Total Topological Charge}\label{1}
\end{table}
 \end{center}
 \subsection{Photon Sphere and Black Holes With Nonlinear Electrodynamics(NLE)}
 The idea that nonlinear electrodynamics might be able to solve the problem of black hole singularities led to many studies to find regular solutions, which led to black holes such as "Bardeen" \cite{55} or "Ayon-Beat-Garcia" \cite{56,57}.
As the last model we chose for our work, we went to a black hole that was selected by "Gao"\cite{58} to investigate the effects of nonlinear electrodynamics and the possibility that a black hole can have more than two event horizons.\\
We study this black hole in three and four-horizon states, and we will show that additional studies confirm the results of the mentioned article in three-horizon states.
\begin{center}
\textbf{Three-Horizons mode }
\end{center}
So, according to \cite{58} for the metric function in the case of three-horizons, we will have
\begin{figure}[H]
 \begin{center}
 \subfigure[]{
 \includegraphics[height=6cm,width=8cm]{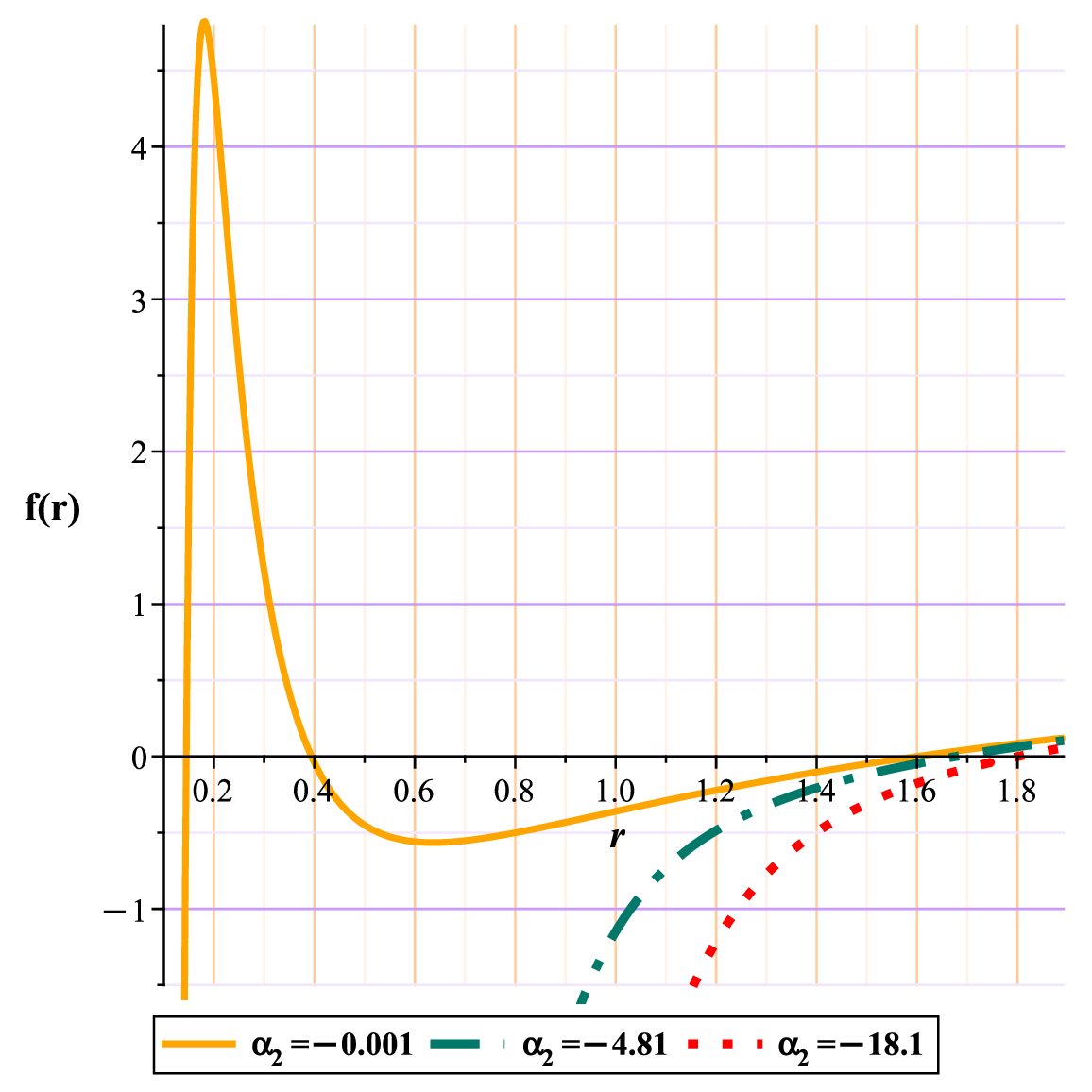}
 \label{8a}}
 
   \caption{\small{Metric function with different $\alpha$ for NLE model  }}
 \label{8}
\end{center}
\end{figure}
\begin{equation}\label{(1)}
f =1-\frac{2 m}{r}+\frac{q^{2}}{r^{2}}+\frac{2 q^{4} \alpha_{2}}{5 r^{6}},
\end{equation}
here $\alpha_{2}$ are dimensional constants,m is mass and q is charge, With respect to(21),(22) and from equations (15) and (16), we will have
\begin{equation}\label{(1)}
H =\frac{\sqrt{25-\frac{50 M}{r}+\frac{25 q^{2}}{r^{2}}+\frac{10 q^{4} \alpha_{2}}{r^{6}}}}{5 \sin \! \left(\theta \right) r},
\end{equation}
\begin{equation}\label{(2)}
\phi^{r}=\frac{\left(15 M \,r^{5}-10 q^{2} r^{4}-5 r^{6}-8 q^{4} \alpha_{2}\right) \sqrt{25-\frac{50 M}{r}+\frac{25 q^{2}}{r^{2}}+\frac{10 q^{4} \alpha_{2}}{r^{6}}}}{5 \sqrt{\frac{-50 M \,r^{5}+25 q^{2} r^{4}+25 r^{6}+10 q^{4} \alpha_{2}}{r^{6}}}\, r^{8} \sin \! \left(\theta \right)},
\end{equation}
\begin{equation}\label{(3)}
\phi^{\theta}=-\frac{\sqrt{25-\frac{50 M}{r}+\frac{25 q^{2}}{r^{2}}+\frac{10 q^{4} \alpha_{2}}{r^{6}}}\, \cos \! \left(\theta \right)}{5 \sin \! \left(\theta \right)^{2} r^{2}}.
\end{equation}
\begin{figure}
 \begin{center}
 \subfigure[]{
 \includegraphics[height=6cm,width=8cm]{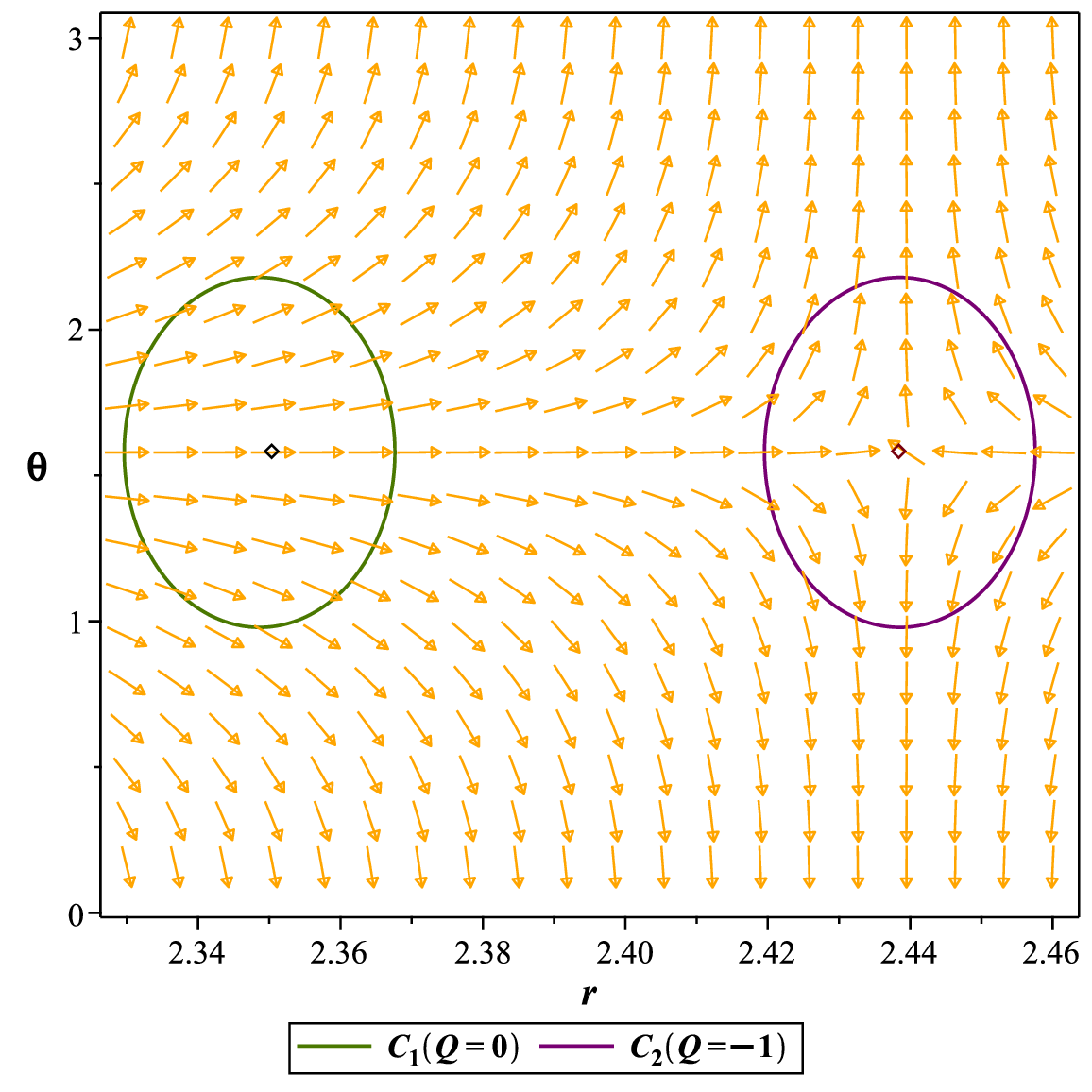}
 \label{6a}}
 \subfigure[]{
 \includegraphics[height=6cm,width=8cm]{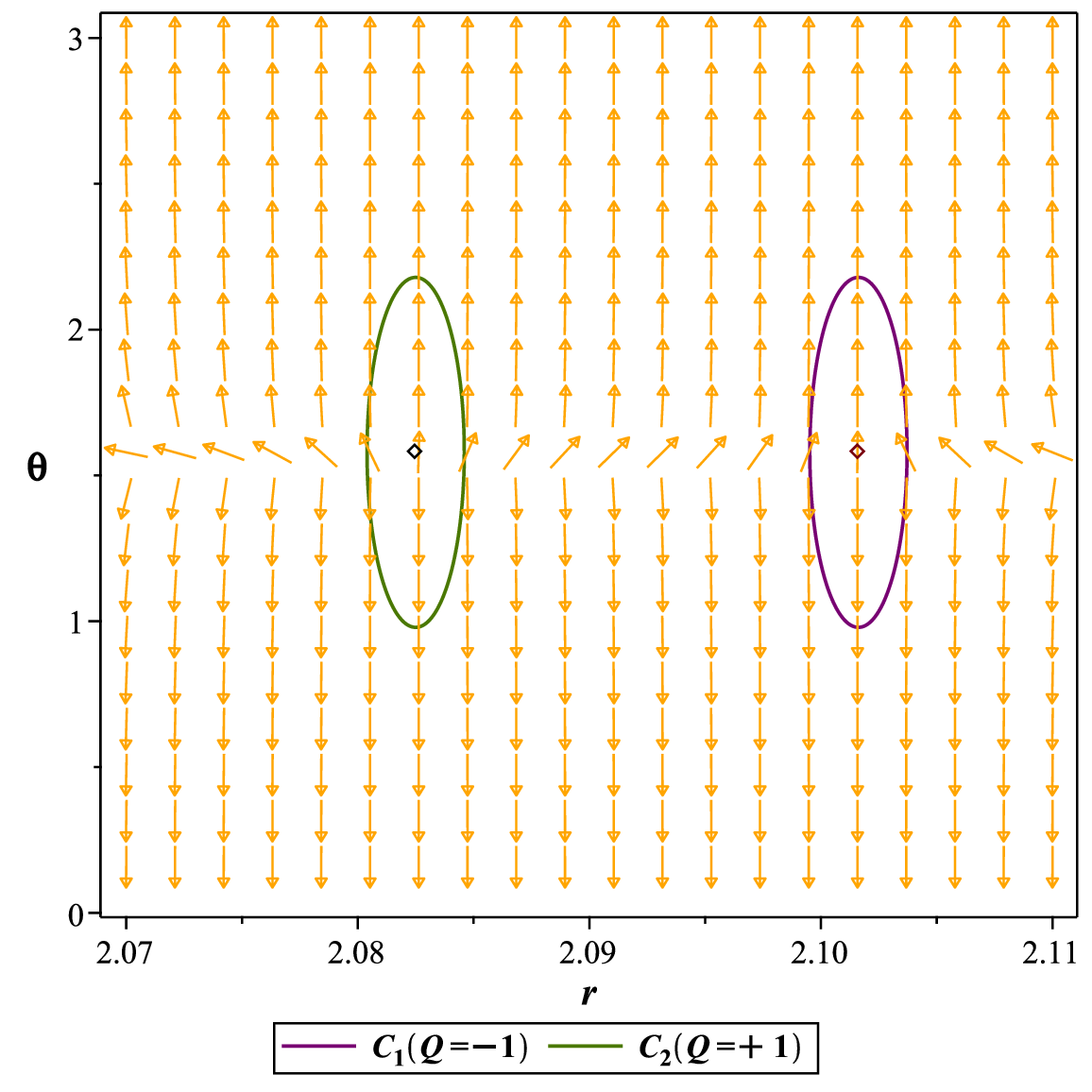}
 \label{6b}}
  \caption{\small{The normal vector field $n$ in the $(r-\theta)$ plane. The photon spheres are located at $ (r,\theta)=(2.438537647,1.57)$ for Fig (15a)with respect to $(q=0.8,m=1,\alpha_{2}=-4.81,)$ and$ (r,\theta)=(2.082498057,1.57)$,$ (r,\theta)=(2.101619064,1.57)$ for Fig (15b) with respect to $(q=0.8,m=1,\alpha_{2}=-18.1,)$}}
 \label{6}
\end{center}
 \end{figure}
The results of "Gao's" work on the black hole showed that to have three horizons, we need $\alpha_{2}<0$ .
It should be stated that our results confirm the above-mentioned limit for black hole behavior.
Our results show that for $\alpha_{2}\leq4.8148$ , the model  shows the behavior of a regular black hole and has an unstable photon sphere ,Figs (6a).
For the range between $4.8148<\alpha_{2}< 18.1064$ the behavior of the model is the naked singularity,Figs (6b) and for a limit greater than that, the system lacks any spherical photon structure, either stable or unstable, and practically, it will most likely lack a black hole structure, table(6). Since the structure of the energy diagrams was similar to the past behaviors, we refrained from displaying them here.
 \begin{center}
\begin{table}[H]
  \centering
\begin{tabular}{|p{3.2cm}|p{4cm}|p{4.5cm}|p{1.5cm}|p{2cm}|}
  \hline
  \centering{NLE black holes(Three-horizons mode)}  & \centering{Fix parametes} &\centering{Conditions}&*TTC&\ $(R_{PLPS})$\\[3mm]
   \hline
  \centering{unstable photon sphere} & $q=0.8,m=1$ & \centering{$\alpha_{2}\leq4.8148$} & $\centering -1$&\ $2.438486243$\\[3mm]
   \hline
  \centering{naked singularity} & $q=0.8,m=1$ & \centering{$4.8148<\alpha_{2}<18.1064$} &\centering $0$&\ $-$ \\[3mm]
   \hline
   \centering{*Unauthorized area} & $q=0.1,m=8$ & \centering{$18.1064<\alpha_{2}$} & $ nothing $&\ $-$ \\[3mm]
   \hline
      \end{tabular}
   \caption{*Unauthorized region: is the region where the roots of $\phi$ equations become negative or imaginary in this region.\\ TTC: *Total Topological Charge\\}\label{1}
\end{table}
 \end{center}
\begin{center}
\textbf{Four-Horizons mode }
\end{center}
Unlike all previous shapes, in this case another parameter has been added to our equations. This means that, since we have only one condition to establish, so our restrictions are increased, and we have to give a value to one of the parameters and add it to our constants table.\\
According to \cite{58} for the metric function in the case of four-horizons, we will have
\begin{equation}\label{(1)}
f \! \left(r \right)=1-\frac{2 m}{r^{2}}+\frac{q^{2}}{r^{2}}+\frac{2 q^{4} \alpha_{2}}{5 r^{6}}+\frac{q^{6} \left(16 \alpha_{2}^{2}-4 \alpha_{3}\right)}{9 r^{10}},
\end{equation}
here $\alpha_{i}$ are dimensional constants,m is mass and q is charge,with respect to(21),(22) and from equations (15) and (16), we will have:
\begin{equation}\label{(2)}
H =\frac{\sqrt{225-\frac{450 m}{r^{2}}+\frac{225 q^{2}}{r^{2}}+\frac{90 q^{4} \alpha_{2}}{r^{6}}+\frac{25 q^{6} \left(16 \alpha_{2}^{2}-4 \alpha_{3}\right)}{r^{10}}}}{15 \sin \! \left(\theta \right) r},
\end{equation}
\begin{equation}\label{(3)}
\phi^{r}=\frac{-15 r^{10}+\left(-30 q^{2}+60 m \right) r^{8}-24 q^{4} \alpha_{2} r^{4}-160 q^{6} \left(\alpha_{2}^{2}-\frac{\alpha_{3}}{4}\right)}{15 r^{12} \sin \! \left(\theta \right)},
\end{equation}
\begin{equation}\label{(4)}
\phi^{\theta}=-\frac{\sqrt{225-\frac{450 m}{r^{2}}+\frac{225 q^{2}}{r^{2}}+\frac{90 q^{4} \alpha_{2}}{r^{6}}+\frac{25 q^{6} \left(16 \alpha_{2}^{2}-4 \alpha_{3}\right)}{r^{10}}}\, \cos \! \left(\theta \right)}{15 \sin \! \left(\theta \right)^{2} r^{2}}.
\end{equation}
\begin{figure}[H]
 \begin{center}
 \subfigure[]{
 \includegraphics[height=6cm,width=8cm]{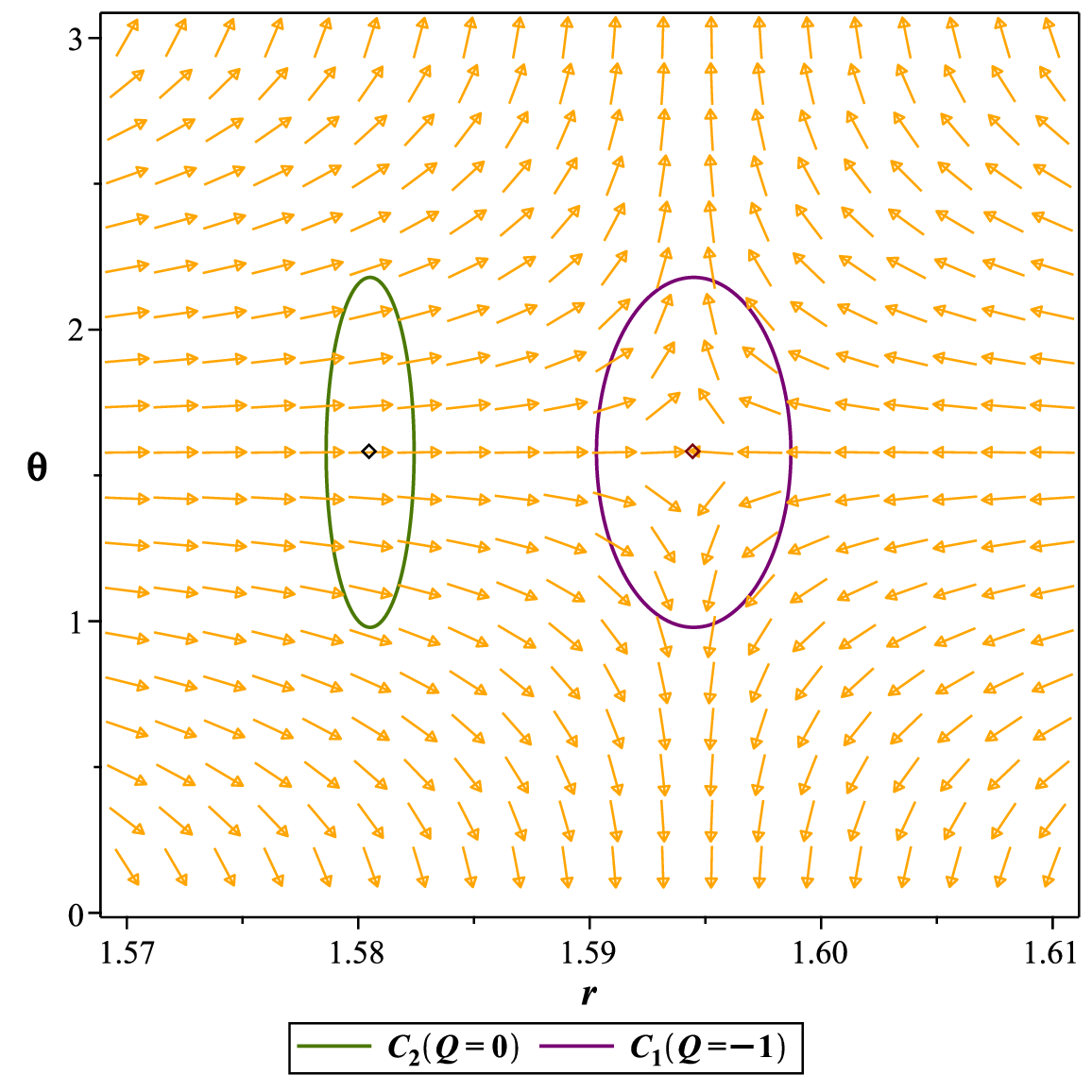}
 \label{7a}}
 \subfigure[]{
 \includegraphics[height=6cm,width=8cm]{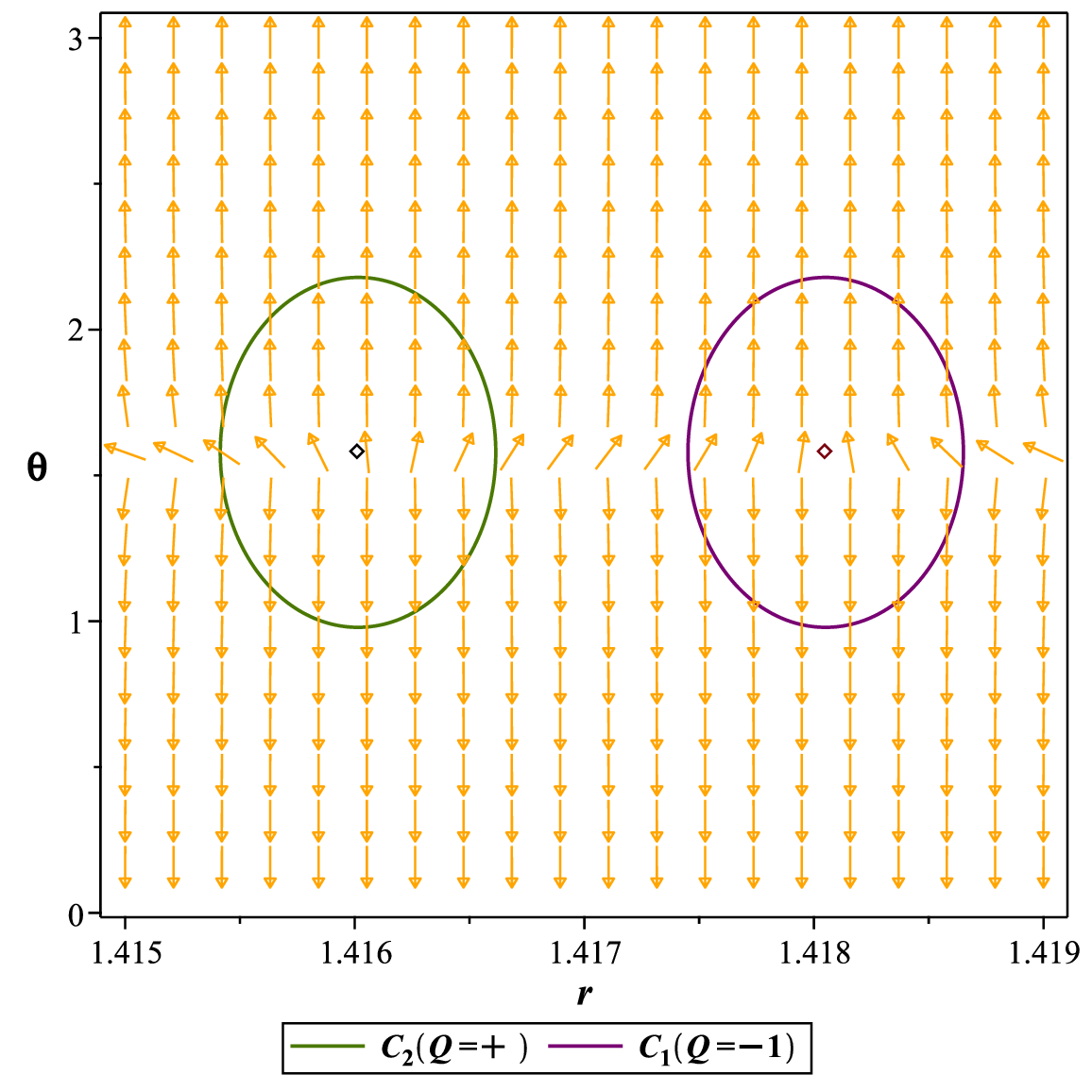}
 \label{7b}}
  \caption{\small{The normal vector field $n$ in the $(r-\theta)$ plane. The photon spheres are located at $ (r,\theta)=(1.594492202,1.57)$ for Fig (16a) with respect to $(q=0.9,m=1,\alpha_{2}=-1,\alpha_{3}=4)$and$ (r,\theta)=(1.416014166,1.57)$,$ (r,\theta)=(1.418051264,1.57)$ for Fig (16b) with respect to $(q=0.9,m=1,\alpha_{2}=-1,\alpha_{3}=-3.2541)$}}
 \label{7}
\end{center}
 \end{figure}
 \begin{center}
\begin{table}[H]
  \centering
\begin{tabular}{|p{3.2cm}|p{4.3cm}|p{4cm}|p{1.5cm}|p{2cm}|}
  \hline
  \centering{NLE black holes(4-horizons mode)}  & \centering{Fix parametes} &\centering{Conditions}& *TTC&\ $(R_{PLPS})$ \\[3mm]
   \hline
    \centering{*Unauthorized area} & $q=0.9,m=1,\alpha_{2}=-1$ & \centering{$\alpha_{3}<-3.2543$} & $ nothing$&\ $-$\\[3mm]
   \hline
  \centering{naked singularity} & $q=0.9,m=1,\alpha_{2}=-1 $& \centering{$-3.2543<\alpha_{3}<3.99$} & $0$&\ $-$\\[3mm]
   \hline
  \centering{unstable photon sphere} & $q=0.9,m=1,\alpha_{2}=-1$ & \centering{$4\leq\alpha_{3}$} & $-1$&\ $1.594492202$ \\[3mm]
   \hline
         \end{tabular}
   \caption{*Unauthorized region: is the region where the roots of $\phi$ equations become negative or imaginary in this region.\\ TTC: *Total Topological Charge}\label{1}
\end{table}
 \end{center}
 \section{Conclusions}
Given the necessity of photon rings and photon spheres for gravitational structures in numerous studies \cite{15,16,17,18,19,20,59,60,61,62,63,64,65,66,67,68,69,70,71}
this article presents a new approach for determining parameter ranges in ultra-compact gravitational objects using topological charges and behavioral patterns associated with photon spheres. To achieve this, we examined models often influenced by PDFM and, in one case, a model with multiple horizons. We meticulously plotted field diagrams, zero points, total topological charges, and effective potential diagrams. Based on these results, we provided a detailed classification of the behavior and characteristics of each model in terms of gravitational influence on photon motion, as shown in Tables 1-7. To better understand the obtained results, it is useful to compare the results presented in the tables before summarizing the overall findings.
As can be seen from the comparison of Tables 1 and 2, in the PDFM black hole, the parameter $\Lambda$ only causes the model to take the form of a black hole within the range of 0 to 2. However, the addition of charge and the AdS radius completely alters this range, such that the model no longer exhibits the form of a naked singularity within the studied values. This behavioral difference, which arises solely from the addition of electric fields to the model, can be quite intriguing.
On the other hand, in the Euler-Heisenberg model, the results appear to be clearer. As shown in Table 3, within a wide range of negative "a" values, the structure takes the form of a naked singularity, and only within a small range of negative "a" values does the structure maintain the form of a black hole. Adding PDFM to the model in this case significantly reduces the range of naked singularity and allows more negative values to keep the model in the form of a black hole. We previously mentioned that negative "a" values could indicate the presence of repulsive interactions that counteract the usual attractive gravitational effects. This means that if the model reaches an extremal or super-extremal state under certain conditions, it would provide suitable conditions for the formation of the Weak Gravity Conjecture (WGC).
It is important to emphasize that although the obtained ranges may not all have physical significance, and to have a fully physical range, they must be interpreted alongside conditions such as energy conditions. However, the mutual influence of parameters and the regions where photon spheres appear can clearly serve as a significant criterion for studying the model's behavior.
This reciprocal relationship between the photon sphere and the gravitational structure is an ideal advantage, and studying it from both perspectives can lead to several interesting outcomes:\\
•  As an initial result, calculating the location of photon spheres, interpreting their topological charge, and determining the effective gravitational range on photon motion (black hole and naked singularity) can serve as a new approach, motivating further study of various models based on this method.
For instance, in this study, we observed that an unstable photon sphere, equivalent to the presence of an unstable maximum in the potential function, only appears in the spacetime influenced by the model when an event horizon exists. Therefore, based on these results and similar experiences \cite{25.1,72,73}, it can be stated that one characteristic of black holes is having an unstable photon sphere. Additionally, the broader spatial coverage of the topological photon sphere study method around the gravitational structure is a key point that distinguishes it from the metric function, meaning that it can be used to determine the gravitational influence range (in the form of black holes + naked singularities) on photon motion.\\
•  One of the main objectives of this article is to classify the parameter space of the studied models based on the existence and topological position of photon and anti-photon spheres. This means that we can precisely determine which parameter range of the model corresponds to a black hole structure or a naked singularity.
For instance, we observed that when the structure is in the form of a naked singularity, one (or more) stable photon spheres always appear alongside an unstable photon sphere in the studied spacetime. In the effective potential representation, this manifests as the appearance of a minimum alongside a maximum.
Therefore, based on the results of this study and similar studies \cite{25.1,72,73}, it can be concluded that if a stable photon sphere is observed within the studied parameter range, this may indicate the possibility of the model existing in the form of a singularity. It is important to emphasize that the meaning of all the concepts discussed so far is not that the obtained ranges necessarily and entirely hold physical meaning. These ranges will be most meaningful when interpreted alongside other pre-existing physical conditions and constraints. Overall, this method can provide a comprehensive view of the behavior of ultra-gravitational structures and can serve as an auxiliary equation, offering better insight into the impact of the effective components in the black hole dynamics of the model.\\
•  Another achievement of this method is that we can adjust the system parameters to force the studied model to exhibit the desired behavior or, conversely, estimate the approximate parameter ranges of the model based on the presence of the photon sphere in a specific region of space. For example, in some gravitational studies, we need a gravitational structure that adheres to specific principles and behaviors. In such cases, where we need the model parameters to fall within a specific range while simultaneously maintaining the structure as a defined black hole, this method can play a significant role. To further clarify, in the Swampland conjecture model study, hypotheses such as the Weak Gravity Conjecture (WGC) require that the studied structure, while being overly charged, still maintains its black hole structure. In this work, we propose that for such scenarios, the topological behavior of the photon sphere can be used, and after adjusting the parameters as desired, the topological photon sphere can be examined to determine whether optimal conditions can be achieved.\cite{72,73,74,75}\\
•  Finally, it is crucial to emphasize that during the formulation of a theory, parameters deemed necessary by the theoretical framework are added. These parameters may conceptually have a defined range from the outset or lack any limitations. In this paper, we have attempted to address the question of what impact these theories, when incorporated into the black hole model action, will have on the behavior of the photon sphere for their permissible values. Additionally, from the perspective of the photon sphere, whether these parameters, with all previous permissible values, will lead to the natural behavior of the studied model or if new constraints need to be applied to these parameters.


\begin{thebibliography}{11}
\bibitem{28}
A Poincaré section is a lower-dimensional subspace of the state space of a continuous dynamical system that intersects a periodic orbit transversally and maps the points of intersection as a discrete dynamical system.
\end{thebibliography}
\end{document}